\documentclass[aps,prb,10pt,twocolumn]{revtex4-2}
\usepackage{amsmath}
\usepackage{graphicx}
\usepackage{amssymb}

\usepackage{gt25}
\usepackage{color} 
\usepackage{graphicx}% Include files
\usepackage{bm}% bold math
\usepackage{amsmath}

\begin{document}
%%%%%%%%%%%%%%%%%%%%%%%%%%
%%%%%%%%%%%%%%%%%%%%%%%%%%%%%%%%%%%%%
\title{
Josephson phase shift and diode effect due to the inverse spin Hall effect
%Anomalous Josephson phase shift and diode effect due to the inverse spin Hall effect
%\\ \textit{or}\\
%Spin-controlled Josephson phase engineering
%\\ \textit{or}\\
%  Spin Control of the Josephson Phase
%  \\ \textit{or}\\
%    Spin-Induced Josephson Phase Shifts
}
\author{Gen Tatara} 
%\email{gen.tatara@riken.jp}
\affiliation{RIKEN Center for Emergent Matter Science (CEMS),
2-1 Hirosawa, Wako, Saitama, 351-0198 Japan}
\author{Yositake Takane}
\affiliation{Graduate School of Advanced Science and Engineering, Hiroshima
University, Higashihiroshima, Hiroshima 739-8530, Japan}
%\affiliation{Graduate School of Advanced Science and Engineering, Hiroshima University, Hiroshima 739-8530, Japan}
\author{Aurelien Manchon}
\affiliation{CINaM, Aix-Marseille University, France}

\date{\today}

\begin{abstract}
We theoretically study the direct and inverse spin Hall effects in a superconductor–normal metal–superconductor junction induced by a spin–orbit interaction that is invariant under spatial inversion. We show that a supercurrent induces a spin Hall effect, leading to a static spin accumulation with opposite polarizations at the two edges, analogous to that in normal conductors. For the inverse effect, we consider a spatially inhomogeneous static magnetic field and show that it induces an anomalous phase shift, which, in the presence of higher harmonics, results in a diode effect.
Unlike Rashba systems studied previously, the present mechanism does not require broken structural inversion symmetry,
%{\color{blue}
since an inhomogeneous magnetic field, equivalent to a spin current, breaks  the inversion symmetry extrinsically.
%}
\end{abstract}

\maketitle

%%%%%%%%%%%%%%%%%%%%%%%%%%%%%%%%%%%%%%%
\renewcommand{\vimp}{v_{\rm imp}}
%%%%%%%%%%%%%%%%%%%%%%%%%%%%%%%%%%%%%%%%%%%%%%%%%%%%%%%%%%%%%%
\section{Introduction}
Spintronic effects in superconductors have attracted considerable interest over the past few decades, driven by the prospect of combining non-dissipative supercurrents with the nonvolatility of spin-based devices.
After pioneering experiments dating back to the late 1970s \cite{Tedrow73}, superconducting spintronics has undergone rapid development over the past decade. Key advances include the realization of long-range spin-triplet supercurrents in ferromagnet–superconductor hybrids \cite{Keizer06,Khaire10,Robinson10,YangSuper10,Jeon21,Sanchez-Manzano22}, electric and magnetic control of superconducting spin transport \cite{Ryazanov01,Jeon23,Jeon26}, and the integration of materials with strong spin–orbit coupling (e.g., topological insulators and Rashba interfaces) to manipulate spin textures \cite{Assouline19,Pal22,Zhang25}. These developments open pathways toward dissipationless memory and logic elements and deepen the interplay between superconductivity and topology \cite{SatoAndo17,Amundsen24}.

Of particular recent interest are systems with broken symmetries \cite{Bergeret16,Daido22,He22,Cayao24,Kokkeler24,Yerin24,Sahoo26}.
In superconductors lacking inversion symmetry, nonreciprocal supercurrent transport becomes possible in the presence of an applied magnetic field or magnetization. This manifests as a direction-dependent critical current, known as the superconducting diode effect, first observed in inversion-asymmetric superlattices \cite{Ando20}.
%%%
The superconducting diode effect has since been demonstrated in Josephson junctions comprising superconductors coupled to normal metals or semiconductors with Rashba spin–orbit coupling \cite{Baumgartner22,Jeon22,Reinhardt24}. Rashba spin-orbit coupling, which arises from the interplay of spin-orbit  interaction and inversion symmetry breaking, generates a uniform spin polarization under an applied current via the Rashba–Edelstein effect, for both dissipative currents and supercurrents \cite{Edelstein90,Edelstein95}.

In superconductor–normal metal–superconductor (SNS) Josephson junctions, Ref.  \cite{Konschelle15} predicted that a supercurrent induces a uniform spin polarization and, conversely, that a magnetic field generates a supercurrent — an effect known as the inverse Rashba–Edelstein effect. This inverse effect was shown to produce an anomalous phase shift between the superconductors. However, such a phase shift alone does not give rise to a diode effect, as it does not break the symmetry of the phase potential and therefore yields identical barrier heights for both current polarities. More recently, it has been shown that this anomalous phase shift can lead to a Josephson diode effect when higher harmonics—arising from multiple interface scattering—are taken into account  \cite{Reinhardt24}.

%%%%%%%%%%%%%%%%
\begin{figure}\centering
\includegraphics[width=0.8\hsize]{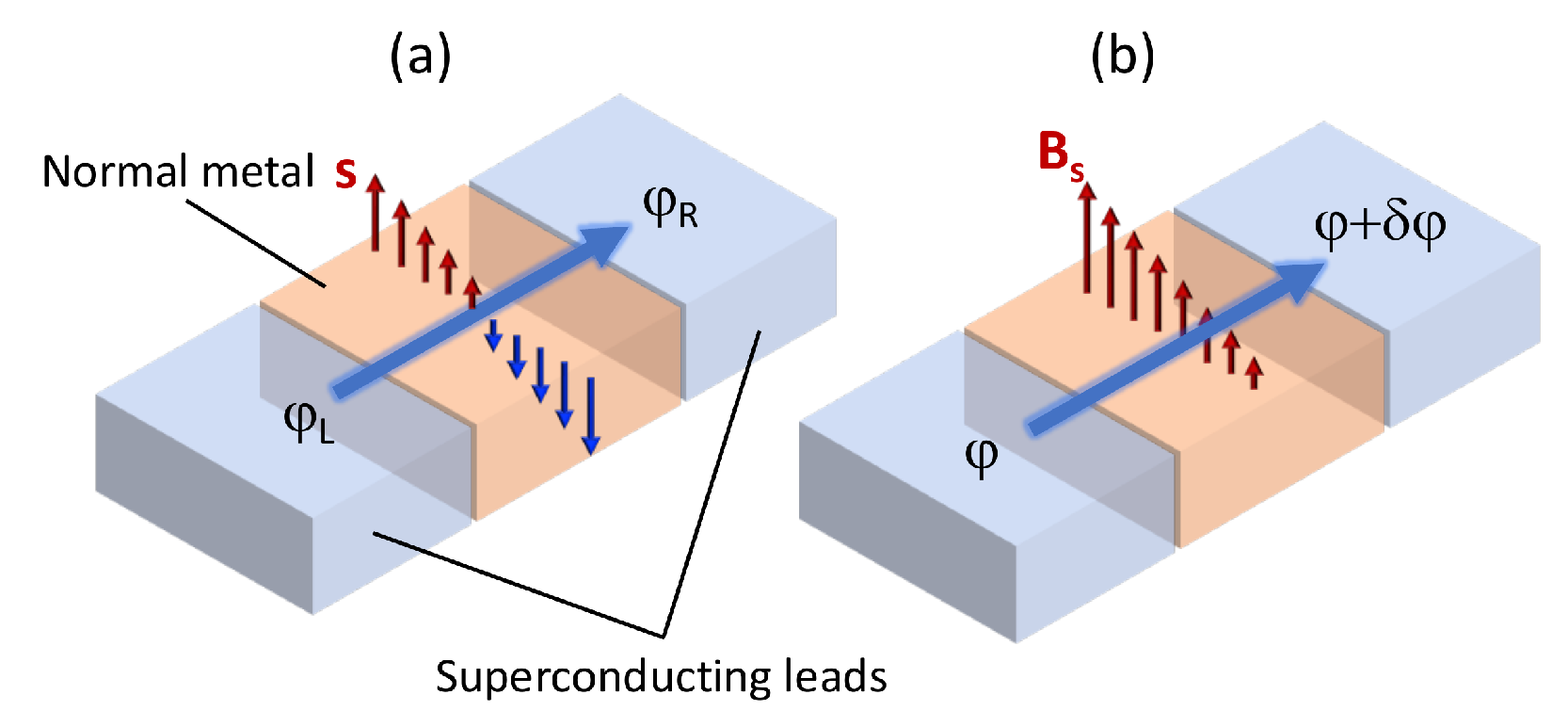}
 \caption{ Schematic figures showing (a) the direct and (b) the inverse spin Hall effect in SNS. In the direct spin Hall effect, equilibrium spin polarization with spatial inhomogeneity  arises as a response to a supercurrent induced by the phase difference.
The inverse effect is induced by a static magnetic field with a spatial gradient, resulting in a supercurrent and an anomalous phase shift.
  \label{FIGsys}}
\end{figure}
%%%%%%%%%%%%%%%%%%%%%%%%%%%%%%%%%%%%%
In this paper, we study  the spin Hall effect (SHE) and its inverse effect (ISHE) theoretically in an SNS junction  (Fig. \ref{FIGsys}) induced by  a spin-orbit interaction that is invariant under spatial inversion.
We show that a supercurrent induces a SHE, leading to a static spin accumulation of opposite polarizations at the two edges, analogous to the behavior in normal conductors. We also investigate the inverse effect by considering a spatially inhomogeneous static magnetic field. Such a magnetic field, acting on the electron spin, generates a steady spin current  \cite{TataraReview19}, which in turn couples to the superconducting channel, inducing an anomalous phase shift.
We show that the anomalous phase shift gives rise to the diode effect by accounting for higher harmonics.
In contrast to the diode effect due to the inverse Rashba-Edelstein effect, the ISHE-induced diode effect does not require the system to lack inversion symmetry, since the spin current breaks inversion symmetry extrinsically.
%{\color{blue}
In our setup, a spatially inhomogeneous magnetic field is essential for the diode effect, as it induces a steady spin current, that breaks  the inversion symmetry.
In the Rashba case with broken structural inversion, in contrast, a homogeneous applied magnetic field is sufficient for the diode effect.
%}

The paper is organized as follows.
We start by summarizing the direct and inverse spin Hall effects in normal metals, as argued in terms of the spin density and electric current in Sec. \ref{SECreciprocal}.
The model of the SNS junction and calculation scheme by use of the lesser Green's function is presented in Sec. \ref{SECmodel}, followed by the derivation of the Josephson current in our scheme in Sec. \ref{SECjj}.
Our study of the supercurrent-induced spin Hall effect starts in Sec. \ref{SECsh}, where we calculate the spin density induced by the Josephson current due to a phase difference at equilibrium,  including to the linear order the self-energy due to each SN interface.
In Sec. \ref{SECish}, the inverse spin Hall effect induced by an applied magnetic field gradient is calculated.
The higher harmonics of the inverse spin Hall current are studied in Sec. \ref{SEChigher} and the Josephson diode effect, the difference in the critical current for superconductivity under  opposite current bias, is studied in Sec. \ref{SECdiode}.
The case of the Rashba spin-orbit interaction, reported previously \cite{Konschelle15}, is briefly argued in Sec. \ref{SECrashba}.
Discussion and conclusion are presented in Sec. \ref{SECsummary}.

%%%%%%%%%%%%%%%%%%%%%%%%%%%%%%%%%%%%%%%%%%%%%%
\section{Premises on spin Hall effect in normal metal \label{SECreciprocal}}
%%%%%%%%%%%%%%%%%%%%%%%%%%%%%%%%%%%%%%%%%%%%%%
Conventionally, the SHE in normal metal is described in terms of the spin current generated by the applied electric field \cite{Hirsch99}.
Since the spin current is not directly measurable, it is converted to an experimentally measurable spin accumulation using the classical spin diffusion equation.
This conventional theory has a fundamental ambiguity arising from the nonconserving nature of spin current in solids.
%%%%%%%%%%%%%%%%%%%%%%%%%%%%%%%%%%%%%%%%%%55
\subsection{Reciprocal relations in terms of spin density}
%%%%%%%%%%%%%%%%%%%%%%%%%%%%%%%%%%%%%%%%%%55
A description free from such a fundamental difficulty can be constructed in terms of the direct response function between spin density and electric current \cite{TataraSH18}.
In the case of a normal metal, the spin density induced by the applied electromagnetic field is written in terms of a vector potential $\Av$ as
\begin{align}
 s_i = \chi_{ij}^{\rm sh} A_j,
\end{align}
where  the response function  $\chi_{ij}^{\rm sh}$ is a correlation function of spin and electric current.
As seen from the symmetry, $\chi_{ij}^{\rm sh}$  does not have a uniform component in  inversion-symmetric systems, and
 is expanded with respect to the external wave vector $\qv$ and frequency $\Omega$ as\cite{TataraSH18}
\begin{align}
  \chi^{\rm sh}_{ij}(\qv,\Omega)
   = i\epsilon_{ijk}q_k (\lambda_0 + i\Omega \lambda_1 +\cdots),
\end{align}
which reads (by use of Maxwell's equation $\dot{\Bv}=-\nabla\times \Ev$)
\begin{align}
 \sv %&= c_0\Bv+c_1 \dot{\Bv}+ \cdots
 = \lambda_0\Bv-\lambda_1 (\nabla\times \Ev) + \cdots,
 \label{sh}
\end{align}
in the ballistic case.
A dynamic term $\lambda_1$ indicates that spin density arises from the vorticity of an electric current, $\nabla\times \jv$, where electric current is $\jv=\sigmab \Ev$, $\sigmab$ being the Boltzmann conductivity.
The SHE corresponds to this term.
In fact, the vorticity arises at the edge of the system, where the current density diminishes, consistent with spin accumulation resulting from the spin Hall effect \cite{TataraSH18}.
Although the effect induces a static spin accumulation in a static electric field, it is a dynamic effect because the electric field induces out-of-equilibrium responses.
Due to Onsager reciprocity, the ISHE induced by an applied magnetic field $ \Bv_{s}$ coupling to the electron spin is written in terms of the same response function as
\begin{align}
j_i = -\chi^{\rm sh}_{ji} \gamma B_{s,j},
\end{align}
where  $\gamma$ is the gyromagnetic ratio.
The current induced by the spin magnetic field thus reads
\begin{align}
 \jv &= - \gamma \lambda_0 \nabla\times  \Bv_{s} - \gamma \lambda_1 \nabla\times  \dot{\Bv}_{s}. \label{ish}
\end{align}
The ISHE, represented by the term $\lambda_1$, is a dynamic effect, while the static effect, $\lambda_0$,  describes  the Amp\`ere's law. The ISHE, being of dynamic origin, is a natural consequence of the fact that normal current reacts to a physical field $\Ev$ and not to the vector potential \cite{Konschelle15}.

%%%%%%%%%%%%%%%%%%%%%%%%%%%%%%%%%%%%%%%%%%%%%%%%%%%%
\subsection{Diffusive case and spin Hall angle}
%%%%%%%%%%%%%%%%%%%%%%%%%%%%%%%%%%%%%%%%%%%%%%%%%%%%
Taking electron diffusion into account,  ballistic behaviors become nonlocal on the scale of the spin diffusion length, $\ell_{\rm s}$, and the spin Hall effect is modified to be \cite{TataraSH18}
\begin{align}
 \sv(\rv)= -\frac{\lambda_{1}}{\sigmab} \int d\rv'D_s(\rv-\rv')(\nabla\times \jv)_{\rv'},
 \label{snormal}
\end{align}
where $D_s(\rv)=\int \frac{dp}{2\pi} \frac{e^{i\pv\cdot\rv}}{(Dp^2+\tau_s^{-1})\tau}$ is the spin diffusion propagator, $D$ being the diffusion constant, and  $\tau_s$ (related to $\ell_{\rm s}$ as $\ell_{\rm s}=\sqrt{D\tau_{\rm s}}$) is the spin relaxation time ($m$ and $\kf$ are the electron mass and the Fermi wavenumber, respectively).
Evaluating the diffusion propagator in one dimension in the $y$ direction perpendicular to the junction, we have
$D_s(y)=\frac{\ell_s}{2\ell^2}e^{-|y|/\ell_s}$, where $\ell$ is the electron mean free path.
Treating the current density without diffusion to vanish suddenly at the system edges, $y=\pm W/2$ ($W$ is the system width), Eq. (\ref{snormal}) leads to
\begin{align}
 s_z(y) =-\lambda_{1} E \frac{\ell_s}{\ell^2} e^{-W/(2\ell_s)} \sinh \frac{y}{\ell_s},
\end{align}
where $j$ is the maximum value of the current along the $x$ (junction) direction.
The maximum spin density at the edge is thus
$s_z(W/2)  \simeq \frac{\ell_s}{2\ell^2}  \lambda_{1}E $.
In terms of the conventional dimensionless spin Hall angle, defined as
$\theta_{\rm sh} =\sigma_{s}/\sigmab$, where $\sigma_{s}$ is the spin Hall  conductivity, the spin accumulation density at the edge is
$s\sim \theta_{\rm sh} \ell_s \dos n E$, $n$ and $\dos$ being the electron density and density of states at the Fermi energy, respectively.
From the two expressions, we see that the spin Hall angle in the conventional picture is
\begin{align}
 \theta_{\rm sh}= \frac{\lambda_{1}}{n\dos \ell^2}.
\end{align}

The coefficient $\lambda_1$  induced by the random spin-orbit interaction was calculated in Ref. \cite{TataraSH18} to be (note that the definitions of the coefficients in Ref. \cite{TataraSH18} are different from here)
\begin{align}
  \lambda_{1} =\frac{\pi}{9}\epsilon_{\rm so}\dos^2 n\ell^2,
\end{align}
where $\epsilon_{\rm so}$ is the energy associated with the spin-orbit interaction.
The spin Hall angle,
\begin{align}
 \theta_{\rm sh}= \frac{\pi}{9}\epsilon_{\rm so} \dos,
 \label{spinhallangle}
\end{align}
 is thus the ratio of the spin-orbit energy and the Fermi energy $\ef(\sim1/\dos)$.

%%%%%%%%%%%%%%%%%%%%%%%%%%%%%%%%%%%%%%%%%%%%%%%%%%%%%%%%%%
\section{Model of SNS junction \label{SECmodel}}
%%%%%%%%%%%%%%%%%%%%%%%%%%%%%%%%%%%%%%%%%%%%%%%%%%%%%%%%%%
We consider an SNS  junction
with a normal metal N with spin-orbit interaction, described by a field Hamiltonian
\begin{align}
 H&= H_{\rm N}+\sum_{\alpha={\rm L},{\rm R}}H_{\alpha} +H_{t},
\end{align}
where  the Hamiltonian for N  represented by field operators $c$ and $c^\dagger$ is
\begin{align}
  H_{\rm N}&=\sum_{\kv\sigma} \ekv c^\dagger_{\kv\sigma}c_{\kv\sigma} +H_{\rm so},
\end{align}
where $\epsilon_{\kv}=\frac{k^2}{2m}-\ef$, $H_{\rm so}$ being the Hamiltonian for the spin-orbit interaction.
The superconductors L and R  are described by
\begin{align}
 H_\alpha &=
 \sum_{\kv}(\begin{array}{c}
{d}_{\alpha\kv\uparrow}^\dagger \\ {d}_{\alpha,-\kv\downarrow}
           \end{array} )
           \lt(\begin{array}{cc}
\epsilon_\kv & \Delta e^{-i\varphi_\alpha}\\
{\Delta} e^{i\varphi_\alpha} & -\epsilon_\kv
           \end{array} \rt)
\lt(\begin{array}{c}
{d}_{\alpha\kv\uparrow} \\ {d}^\dagger_{\alpha,-\kv\downarrow}
           \end{array} \rt),
\end{align}
where ${d}_{\alpha\kv\sigma}^\dagger $ and ${d}_{\alpha\kv\sigma} $ are field operators of the electron in superconductor $\alpha=$L,R with wave vector $\kv$ and spin $\sigma$.
The magnitude of the order parameter is $\Delta$, and its phase is $\varphi_\alpha$.
The interface hopping is represented by
\begin{align}
 H_t &= \sum_{\alpha,\sigma=\uparrow,\downarrow} \sum_{\rv\in {{\rm I}_\alpha}} t[  d_{\alpha\sigma}^\dagger c_{\sigma}
    +  c_{\sigma}^\dagger d_{\alpha\sigma}],
\end{align}
where $t$ is the hopping amplitude  assumed to be real and spin-independent.
The hopping is treated as occurring at the interface, ${\rm I}_\alpha$. %, which we shall approximate as point-like.
Although there have been extensive theoretical works on SNS junctions \cite{Golubov04},  our model is minimal and sufficient to capture the essence of supercurrent-driven spintronics.

%%%%%%%%%%%%%%%%%%%%%%%%%%%%%%%%%%%%%%%%%%%%%%%%%%%%
\subsection{The Green's function in N}
%%%%%%%%%%%%%%%%%%%%%%%%%%%%%%%%%%%%%%
We employ the path-ordered (Keldysh) Green's function method, as physical observables are presented in  a more transparent way than in the thermal Green's functions (as in \cite{Kresin86}), although the two approaches give the same result for equilibrium properties.
Physical quantities in N are calculated by use of the lesser Green's function in N,
$G^{<}=i\average{c^\dagger c}$.
The effects of two superconductors, L and R, are included perturbatively  as self-energies
due to the L- and R-superconductors assuming a weak junction \cite{Kresin86}.
A dissipative channel described by electron-hole propagation,
arising from the normal Green's function in the superconductor, is irrelevant and is neglected.
 The supercurrent channel is induced by the superconducting component of the self-energy,
 whose advanced component is
 $\Sigma_{\alpha}^\adv=t^2\sum_{\kv}{F}_{\alpha\kv}^\adv$, where
 \begin{align}
 {F}_{\alpha\kv}^\adv(t,t')
 = i\theta(t'-t)\average{ \{{d}_{\alpha,\kv,\uparrow}(t) ,{d}_{\alpha,\kv,\downarrow}(t')\} },
 \end{align}
is the advanced component of  the anomalous Green's function.
 Here, the summation over the wave vector is taken independently  of N approximating the interface $I_\alpha$ as pointlike.
 The conjugate self-energy is
  \begin{align}
   \overline\Sigma_{\alpha}^\adv=t^2\sum_{\kv}\overline{F}_{\alpha\kv}^\adv,
  \end{align}
   where
 \begin{align}
 \overline{F}_{\alpha\kv}^\adv(t,t')
 = i\theta(t'-t)\average{ \{{d}^\dagger_{\alpha,\kv,\uparrow}(t) ,{d}^\dagger_{\alpha,\kv,\downarrow}(t')\} }.
 \end{align}

 %%%%%
 %%%%%%%%%%%%%%%%
\begin{figure}\centering
\includegraphics[width=0.7\hsize]{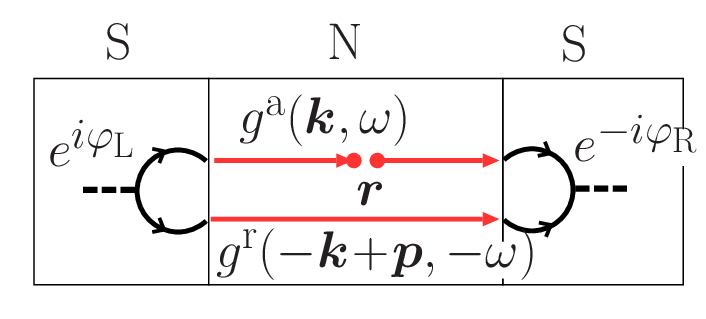}
 \caption{ Schematic diagram showing the
 lesser Green's function in N, $G^{<}  (\rv,\rv,\omega)$ (the Fourier transform of
 $G^{<}  (\rv,t,\rv',t')$ at equal spatial positions, $\rv=\rv'$), to first order in the self-energies from L and R.
  \label{FIGSNS}}
\end{figure}
%%%%%%%%%%%%%%%%%%%%%%%%%%%%%%%%%%%%%
The Green's function contribution representing the second order Josephson coupling in the interface hopping is (suppressing the spin index) (Fig. \ref{FIGSNS})
\begin{align}
 G^{<} & (\rv,t,\rv',t')
  = \int_{-\infty}^{\infty}\! dt_1\! \int\! dt_2\! \int\! dt_3 \! \int\!dt_4 \sum_{\Rv_L,\Rv_R}
  [g(\rv,t,\Rv_L,t_1)
  \nnr & \times
  \Sigma_{\rm L}(t_1,t_2) g(\Rv_R,t_3,\Rv_L,t_2)
  \bar{\Sigma}_{\rm R}(t_3,t_4) g(\Rv_R,t_4,\rv',t')]^<
  \nnr &
+({\rm L}\leftrightarrow{\rm R}),
\label{GlessN}
\end{align}
where $g$ denotes the Green's functions including the spin-orbit interaction, and without superconductors, $^<$ denotes the lesser component, and $\Rv_{\alpha}$ denotes the positions at the interface.
%As $\Sigma_{\alpha}^\adv\propto e^{i\varphi_\alpha}$ and $\overline\Sigma_{\alpha}^\adv\propto e^{-i\varphi_\alpha}$, Eq. (\ref{GlessN}) is a function of phase difference, $\varphi$.
Evaluating the lesser component at equlibrium and at equal times, we obtain
\begin{align}
 G^{<}& (\rv,t,\rv',t)
=
\sumom f(\omega)
\nnr &
 g^\adv(\rv-\Rv_L,\omega) \Sigma_{\rm L}^\adv(\omega)  g^\ret(\Rv_R-\Rv_L,-\omega) \bar{\Sigma}^\adv_{\rm R}(\omega) g^\adv(\Rv_R-\rv',\omega)
 \nnr &
 - (\ret\leftrightarrow\adv)
+({\rm L}\leftrightarrow{\rm R}),
 \label{GNLR0}
\end{align}
where $f(\omega)=[e^{\beta\omega}+1]^{-1}$ is the Fermi-Dirac distribution function, $\beta=1/(\kb T)$ being the inverse temperature.
Note that due to an anomalous time ordering as a result of the anomalous particle-particle propagation, one of the Green's functions has a frequency with opposite sign and retarded and advanced interchanged.
The anomalous propagation thus leads to a contribution of a product of the retarded and advanced Green's functions, which  guarantees the long-ranged supercurrent.

%%%%%%%%%%%%%%%%%%%%%
\subsection{Self-energy due to S}
%%%%%%%%%%%%%%%%%%%%%
The anomalous self-energies are written using the phase, $\varphi_{\rm L}$ and $\varphi_{\rm R}$, as
\begin{align}
\Sigma_{\rm L}^\adv &= \Sigma^\adv e^{-i\varphi_{\rm L}}, &
\bar\Sigma_{\rm R}^\adv &= \Sigma^\adv e^{i\varphi_{\rm R}},
\end{align}
where
\begin{align}
\Sigma^\adv(\omega) &= t^2 \Delta \sum_{\kv}\frac{1}{2\xi_\kv}  \sum_{\pm}(\pm)\frac{1}{\omega\pm \xi_{\kv}-i0},
%=t^2 \Delta \frac{\pi\dos}{\sqrt{\Delta^2-\omega^2}}\equiv \Sigma
\end{align}
with $\xi_{\kv}=\sqrt{\ekv^2+\Delta^2}$, is common for L and R.
%in the absence of a current  bias.
Using an approximation that the density of states $\dos$ is a constant and at small $\omega(\leq \Delta)$, we obtain
\begin{align}
\Sigma^\adv(\omega) &= t^2 \Delta \frac{\pi\dos}{\sqrt{\Delta^2-\omega^2}}
\equiv \Sigma(\omega).
\end{align}

%%%%%%%%%%%%%%%%%%%%%%%%%
\section{Josephson current in the diffusive case \label{SECjj}}
%%%%%%%%%%%%%%%%%%%%%%%%%
%%%%%%%%%%%%%%%%%%%%%%%%%%%%%%%%%%%%%%%%%%%%%%%%%%%%%%%%%%%%%%

%%%%%%%%%%%%%%%%
\begin{figure}\centering
\includegraphics[width=0.45\hsize]{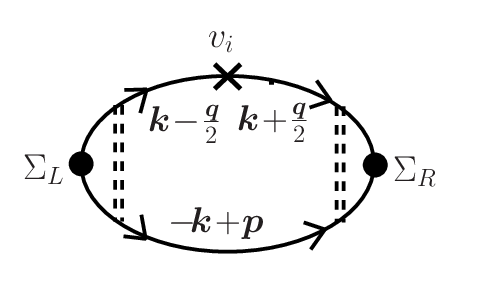}
\includegraphics[width=0.45\hsize]{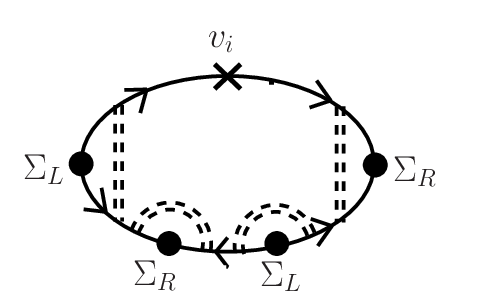}
 \caption{Diagrams representing the Josephson current in the disordered case.
 A pair of double dotted lines denotes particle-particle propagation (cooperon).
 $v_i$ denotes a current vertex with
 %{\color{magenta}
 an external wave vector $\qv$, which represents the spatial variation of the current.
 The wave vector $\pv$ is the total wave vector of the propagating electron pair, i.e., the cooperon wave vector.
 %}
 Left: Contribution in second order in the self-energy due to superconductors.
 Right: Fourth-order contribution.
  \label{FIGjjdif}}
\end{figure}
%%%%%%%%%%%%%%%%%%%%%%%%%%%%%%%%%%%%%
Before exploring the spin Hall effect, let us first reproduce the Josephson current within the present formalism.
The derivation is essentially equivalent to that in Ref. \cite{Kresin86}, except that the latter is based on the thermal Green's function formalism.

In the diffusive case, the Josephson current is supported by long-range particle-particle propagation called the cooperon.
The current to second order in the self-energy reads (Fig. \ref{FIGjjdif})
\begin{align}
 j_{\rm J}
& =
i\sumom (2f(\omega)-1) [\Sigma(\omega)]^2 \sum_{\pv}
(e^{i\varphi} e^{i\pv\cdot\Lv} + e^{-i\varphi}  e^{-i\pv\cdot\Lv})
\nnr & \times
K_{\rm J}(\pv,\omega)(\nimp\vimp^2\Pi_{\pv,\omega})^2 D(\pv,\omega)^2,
 \label{JJDif1}
\end{align}
where $\nimp$ and $\vimp$ are the concentration and strength of impurity potential, respectively, and
$\Lv\equiv(L,0,0)$,  $L$ being the distance between the two interfaces.
The electron pair propagation is represented by an amplitude
\begin{align}
 \Pi_{\pv,\omega} &\equiv \sum_{\kv} g^\adv_{\kv,\omega} g^{\ret}_{-\kv+\pv,-\omega},
\end{align}
and
\begin{align}
D({\pv,\omega}) &=\frac{1}{\tau} \frac{1}{Dp^2-2i\omega}
\end{align}
is the cooperon propagator with  the wave vector $\pv$, $\tau$ being the elastic lifetime.
The bare response  function is
\begin{align}
K_{\rm J}(\pv,\omega) =\sum_{\kv}
\tr[{g^\adv_{\kv,\omega}}\hat{v}_x {g^\adv_{\kv,\omega}}{g^\ret_{-\kv+\pv,-\omega}} ],
\end{align}
$\hat{v}_i$ ($i=x,y,z$) being the velocity operator.
Expanding with respect to $\pv$, we obtain
\begin{align}
K_{\rm J}(\pv,\omega) = -4\pi \dos \tau^2 Dp_x.
\end{align}
In the diffusive case, $ \Pi_{\pv,\omega}\simeq 2\pi \dos \tau$, and thus $\nimp\vimp^2 \Pi_{\pv,\omega}\simeq 1$.
We carry out the summation over $\pv$ in three dimensions.
The cooperon propagator including, a factor $p_x$ arising from $K_{\rm J}$, is calculated as
\begin{align}
\sum_{\pv}
 \frac{p_x e^{\pm i\pv\cdot\Lv}}{(Dp^2-2i\omega)^2}
 &= \frac{\pm iL}{2D} D_0(L,\omega).
\end{align}
The amplitude $D_0$ is
\begin{align}
 D_0(L,\omega) &\equiv
\sum_{\pv}
 \frac{e^{i\pv \cdot\Lv}}{Dp^2-2i\omega}
 \nnr
 &= \frac{1}{4\pi LD} e^{-\sqrt{\frac{|\omega|}{D}}L}
   e^{\pm i \sqrt{\frac{|\omega|}{D}}L},
\end{align}
where  sign corresponds to $\omega>0$ and $\omega<0$.
We focus on the low-frequency regime, $|\omega|\ll\Delta$ and approximate $\Sigma(\omega)$ as a constant $\Sigma$.
Integration over the frequency then reads
\begin{align}
 \sumom(2f(\omega)-1)\sum_{\pv}
 \frac{p_x e^{\pm i\pv\cdot\Lv}}{(Dp^2-2i\omega)^2}
   &= \mp\frac{L}{4\pi^2 DL^3} F(L/\ell_T),
\end{align}
where
\begin{align}
\ell_T\equiv \sqrt{\frac{D}{T}},
\end{align}
is thermal diffusion length, and
\begin{align}
 F(\mu) & \equiv \int_0^\infty dx \tanh \lt(\frac{x^2}{2\mu^2} \rt) e^{-x} \sin x.
 \label{Fmudef}
\end{align}
The diffusive Josephson current, including the self-energy to second order, therefore, is
\begin{align}
 j_{\rm J}
  &
  = j_{c2}\sin\varphi,
  \label{Jcurrentuniform}
\end{align}
with the critical current at the second order
\begin{align}
j_{c2} &=  -\frac{2}{\pi} \frac{\dos \Sigma^2 }{L^2} F(L/\ell_T).
\end{align}

%%%%%%%%%%%%%%%%%%%%%%%%%%%%%%%%%%%%%%%%
\subsection{Spatially-resolved Josephson current}
%%%%%%%%%%%%%%%%%%%%%%%%%%%%%%%%%%%%%%%%
%{\color{magenta}
To study the spin Hall effect, it is necessary to investigate the spatial dependence of the Josephson current by calculating its $\qv$-dependence, where $\qv$ denotes the wave vector in the Fourier transform of the Josephson current.
%}
Including finite wave vector $\qv$ at the current vertex, the current in the diffusive regime in Fig.  \ref{FIGjjdif} reads
\begin{align}
 j_{\rm J}(\qv)
& =
i\sumom (2f(\omega)-1) [\Sigma(\omega)]^2 \sum_{\pv}
(e^{i\varphi} e^{i\pv\cdot\Lv} + e^{-i\varphi}  e^{-i\pv\cdot\Lv})
\nnr & \times
\frac{K_{\rm J}(\pv,\omega)}{\tau^2}
\frac{1}{D\lt(p+\frac{\qv}{2}\rt)^2-2i\omega}
\frac{1}{D\lt(p-\frac{\qv}{2}\rt)^2-2i\omega}.
 \label{JJDifq1}
\end{align}
Defining a function which determines the spatial dependence of the current as
\begin{align}
 \Gamma_j(\qv,\omega)
& \equiv
 \sum_{\pv}
 p_j e^{i\pv\cdot\Lv}
\frac{1}{D\lt(p+\frac{\qv}{2}\rt)^2-2i\omega}
\frac{1}{D\lt(p-\frac{\qv}{2}\rt)^2-2i\omega},
 \label{PiDif}
\end{align}
we have
\begin{align}
 j_{\rm J}(\rv)
& = 8\pi \dos D
\sumom (2f(\omega)-1) [\Sigma(\omega)]^2
 \Gamma_x(\rv,\omega) \sin\varphi.
 \label{JJDifq2}
\end{align}
%The uniform component of the current is the one with $\qv=0$ in (\ref{JJDifq1}), which reduces to Eq. (\ref{Jcurrentuniform}).

%%%%%%%%%%%%%%%%
\begin{figure}\centering
\includegraphics[width=0.7\hsize]{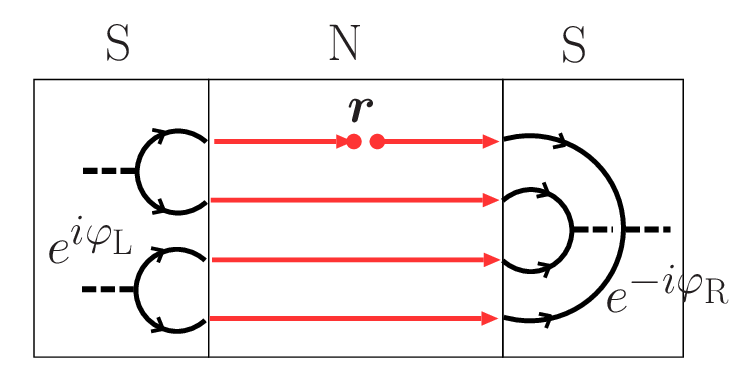}
 \caption{Diagram representing the  fourth-order supercurrent, which is carried by the two particle-particle propagators flowing in the same direction. Other diagrams with particle-hole propagation do not contribute to the supercurrent channel.
  \label{FIG4thGf}}
\end{figure}
%%%%%%%%%%%%%%%%%%%%%%%%%%%%%%%%%%%%%
%%%%%%%%%%%%%%%%%%%%%%%%%%%%%%%%%%%%%%%%
\subsection{Fourth-order Josephson current}
%%%%%%%%%%%%%%%%%%%%%%%%%%%%%%%%%%%%%%%%
The Josephson current, including the self-energy to fourth order,  is necessary to study the diode effect.
In terms of the Green's function, $G^<$, the fourth-order supercurrent is represented by the diagram with two particle-particle propagators as shown in Fig. \ref{FIG4thGf}.
Diagrams containing the particle-hole propagation contribute to normal current and are not considered.
In the diffusive case, it includes four cooperons modifying each of the four vertices representing the hopping to the superconductor as in Fig. \ref{FIGjjdif}.
Insertion of more cooperons is possible, but the contribution is small due to the narrow phase space.
The dominant fourth-order Josephson current is, therefore,
\begin{align}
 j_{\rm J}^{(4)}
& =
i\sumom (2f(\omega)-1) [\Sigma(\omega)]^4 \sum_{\pv}
(e^{2i\varphi} e^{i\pv\cdot\Lv} + e^{-2i\varphi}  e^{-i\pv\cdot\Lv})
\nnr & \times
K^{(4)}_{\rm J}(\pv,\omega)(\nimp\vimp^2\Pi_{\pv,\omega})^4 D(\pv,\omega)^4,
 \label{JJDif4}
\end{align}
with
\begin{align}
K_{\rm J}^{(4)}(\pv,\omega) &= \sum_{\kv}
\frac{k_x}{m} ({g^\adv_{\kv,\omega}})^3 ({g^\ret_{-\kv+\pv,-\omega}} )^2
 \simeq %24 \pi \dos D \tau^4 p_i=
-6\tau^2\times K_{\rm J}(\pv,\omega).
\end{align}
The cooperon contribution is
\begin{align}
\sum_{\pv} \frac{p_x e^{i\pv\cdot\Lv}}{(Dp^2-2i\omega)^4}
&=
\frac{iL}{6D}
\sum_{\pv} \frac{e^{i\pv\cdot\Lv}}{(Dp^2-2i\omega)^3}.
\end{align}
The summation in three dimensions is
\begin{align}
\sum_{\pv} &\frac{e^{i\pv\cdot\Lv}}{(Dp^2-2i\omega)^3}
=
\frac{1}{2^7 \pi D^{3/2}}\frac{1}{|\omega|^{3/2}}
\nnr &\times
 \lt[ \pm i\lt(2L\lt(\frac{|\omega|}{D}\rt)^{1/2}+1 \rt)-1\rt] e^{-\sqrt{\frac{|\omega|}{D}}L(1\mp i)},
\end{align}
and we have
\begin{align}
 \sumom(2f_\omega-1)\sum_{\pv}
 \frac{p_x e^{i\pv\cdot\Lv}}{(Dp^2-2i\omega)^4}
   &=-\frac{L^2}{3\cdot2^7 \pi^2D^3} F_4(L/\ell_T),
\end{align}
where
\begin{align}
 F_4(\mu) & \equiv \int_0^\infty \frac{dx}{x^2} \tanh \lt(\frac{x^2}{2\mu^2} \rt) {e^{-x}}
 \lt[ (2x+1)\cos x - \sin x \rt].
\end{align}
The fourth order Josephson current is, therefore,
\begin{align}
 j_{\rm J} ^{(4)}
  &= \frac{1}{8\pi} \dos \Sigma^2 \frac{\Sigma^2L^2 }{D^2} F_4(L/\ell_T) \sin 2\varphi.
\end{align}
Josephson current to fourth order of the self-energy in the diffusive regime is
\begin{align}
 j_{\rm J}
  &= j_{c2} \sin \varphi+ j_{c4} \sin 2\varphi,
\end{align}
where
\begin{align}
 j_{c2} &= -\frac{2}{\pi} \frac{\dos \Sigma^2 }{L^2} F(L/\ell_T),
\nnr
 j_{c4} &= \frac{1}{8\pi} \dos \Sigma^2 \frac{\Sigma^2L^2 }{D^2} F_4(L/\ell_T).
\end{align}
The ratio of the two, which is crucial for the magnitude of the Josephson diode effect, is $j_{c4} / j_{c2}= -\frac{1}{16\pi} \lt(\frac{\Sigma L^2 }{D}\rt)^2 \frac{F_4}{F}$, and is governed by a parameter
$\frac{\Sigma L^2 }{D}= 3\Sigma \tau (L/\ell)^2$.

%%%%%%%%%%%%%

%%%%%%%%%%%%%
\section{Supercurrent induced spin Hall effect \label{SECsh}}
%%%%%%%%%%%%%
We here explore these responses for the supercurrent in an SNS junction.
The normal metal (N) has a  spin-orbit interaction arising from random impurity,  represented by a field Hamiltonian
\begin{align}
 H_{\rm so} &= i\lambdaso \vimp\sum_{\kv\kv'}\sum_{i}e^{-i(\kv'-\kv)\cdot\Rv_i} c^\dagger_{\kv'}[(\kv'\times\kv)\cdot\sigmav]c_\kv,
\end{align}
where $\lambdaso$ is a constant,
% $\vimp$ is the strength of the impurity potential,
$\Rv_i$ is the position of impurities and $i$ is the label of impurities.
%{ \color{magenta}
This interaction for a fixed configuration of impurities breaks translational and space-inversion invariance, since $e^{-i(\kv'-\kv)\cdot\Rv_i}\ra e^{i(\kv'-\kv)\cdot\Rv_i}$ under spatial inversion.
For a physically meaningful momentum-conserving result, average over random impurity positions needs to be taken \cite{ColemanIMBP15}.
After the averaging, the inversion and translational invariances are restored.
Our model therefore  does not break inversion symmetry, unlike the Rashba spin-orbit case.
We shall demonstrate that even in this case, without structurally broken inversion, the Josephson diode effect emerges, since the application of a magnetic field gradient (equilibrium spin current) breaks the symmetry.
%}

The supercurrent  induced by the two superconductors is carried by the spin-singlet electron-electron pair propagation, and thus,
the SHE is represented by a nonlocal diagram  connecting L and R superconductors, instead of a local vector potential as in the normal metal case.

%{\color{magenta}
The inhomogeneous spin density induced by the spin Hall effect is studied using a gradient expansion. We consider the Fourier transform $\sv(\qv)$ of the spin density, where $\qv$ is the wave vector, and expand the result to the linear order of $\qv$.
%}
%%%%%%%%%%%%%%%%
\begin{figure}\centering
\includegraphics[width=0.45\hsize]{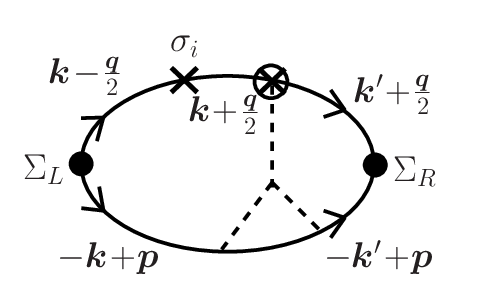}
\includegraphics[width=0.45\hsize]{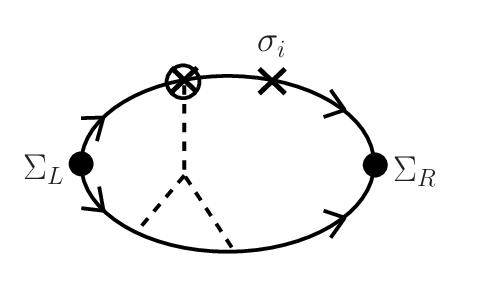}
 \caption{ Diagrams contributing equally to the spin accumulation induced by the SHE induced by random spin-orbit interaction (denoted by $\otimes$) due to impurity (dotted line) in the ballistic case.
  The spin vertex is denoted by $\times$, and black circles denote self-energies $\Sigma_{\rm L}$ and $\Sigma_{\rm R}$ of the  L and R superconductors. The external wave vector is $\qv$, and $\pv$ is the cooperon wave vector.
  \label{FIGshSNS}}
\end{figure}
%%%%%%%%%%%%%%%%%%%%%%%%%%%%%%%%%%%%%

%%%%%%%%%%%%%%%%
\begin{figure}\centering
\includegraphics[width=0.4\hsize]{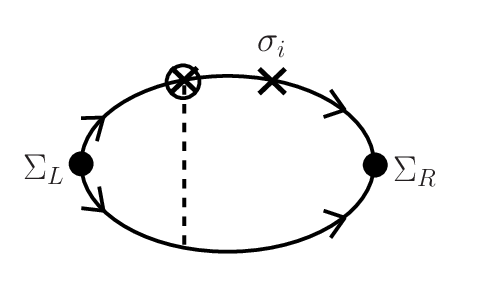}
 \raisebox{2\baselineskip}{\Large{$=$}}
\includegraphics[width=0.4\hsize]{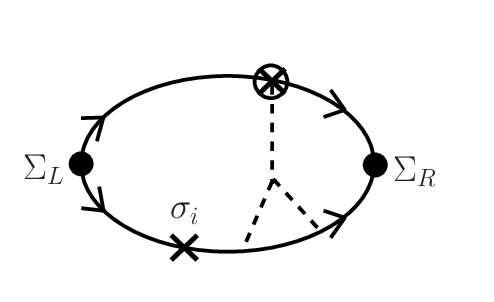}
 \raisebox{2\baselineskip}{\Large{$=0$}}
 \caption{ Diagrams that do not contribute to the supercurrent-induced spin Hall effect.
  \label{FIGshSNSzero}}
\end{figure}
%%%%%%%%%%%%%%%%%%%%%%%%%%%%%%%%%%%%%

%%%%%%%%%%%%%%%%%%%%%%%%%%%%%%%%%%%%%%%%%%%%%%%%%%%%%%%%%%%%%%%
\subsection{Ballistic case}
%%%%%%%%%%%%%%%%%%%%%%%%%%%%%%%%%%%%%%%%%%%%%%%%%%%%%%%%%%%%%%%
In the ballistic case, where electron diffusion is neglected, the spin accumulation induced by the supercurrent in N is
\begin{align}
 s_{i}(\qv)
& =
i\sumom (2f(\omega)-1) [\Sigma(\omega)]^2\sum_{\pv}
(e^{i\varphi}e^{i\pv\cdot\Lv}  +
e^{-i\varphi}e^{-i\pv\cdot\Lv}) \nnr & \times
K_{s,i}(\qv,\pv,\omega),
 \label{spin1}
\end{align}
where $K_{s,i}$ is  the spin Hall  response function, shown diagrammatically in Fig. \ref{FIGshSNS},
\begin{align}
 K_{s,i}& (\qv,\pv,\omega) =
  i\lambdaso\nimp\vimp^3
 \sum_{\kv\kv'\kv''}
 \lt[\lt(\kv'+\frac{\qv}{2}\rt)\times\lt(\kv+\frac{\qv}{2}\rt)\rt]_i
 \nnr & \times
  \tr[ g_{\kv'+\frac{\qv}{2}}^\adv(\omega)g_{\kv+\frac{\qv}{2}}^\adv(\omega) \sigma_i  g_{\kv-\frac{\qv}{2}}^\adv(\omega) \nnr & \times
  g_{-\kv'+\pv}^\ret (-\omega)g_{-\kv+\pv}^\ret (-\omega)g_{\kv''}^\ret (-\omega)].
  \end{align}
As in the anomalous Hall and spin Hall effects in normal metal \cite{Dugaev01,TataraSH18}, the third order of the impurity potential is dominant, while the second order contribution vanishes, being purely imaginary (Fig. \ref{FIGshSNSzero}).
The contribution with the spin vertex and the spin-orbit interaction acting on different sides of the bubble, Fig. \ref{FIGshSNSzero}, vanishes, too.
The upside-down diagram, i.e., the retarded and advanced interchanged, is $\omega\ra-\omega$ contribution, resulting in a term with hole nature, $f(-\omega)=f(\omega)-1$.
Expanding with respect to the external wave vector $\qv$ to the linear order, we obtain
\begin{align}
 K_{s,i}&(\qv,\pv,\omega) =
  \lambdaso\pi \dos\nimp\vimp^3
 \sum_{\kv\kv'}\frac{q_j}{2}
 \nnr &\times
 [(\kv'\times \kv)_i (\partial_{k'_j}   g_{\kv'}^\adv(\omega))
 +
 \epsilon_{ijk}(\kv-\kv')_k  g_{\kv'}^\adv(\omega)
 ] \nnr & \times(g_{\kv}^\adv(\omega))^2 g_{-\kv'+\pv}^\ret (-\omega)g_{-\kv+\pv}^\ret (-\omega).
  \end{align}
Evaluating the summation over wave vectors in the case of small $\pv$ and $\omega$, it is
\begin{align}
K_{s,i}(\qv,\pv,\omega)
&=
  -5\pi^2 \lambdaso  \dos^2\vimp mD \tau^2(\qv\times \pv)_i.
  \end{align}
The response function odd in $\pv$ means that the induced spin is proportional to $\sin\varphi$ and to the the Josephson current.
The factor $\qv$ indicates that the spin accumulation is proportional to the vorticity of the Josephson current induced at the edge.
%{\color{magenta}
In the present formalism, the wave vector $\pv$ is provided by the SN interfaces, while $\qv$ for the spin density profile is provided by the random impurities.
%}

%%%%%%%%%%%%%%%%%%%%%%%%%%%%%%%%%%%%%%%%%%%
\subsection{Diffusive case}
\begin{figure}\centering
\includegraphics[width=0.6\hsize]{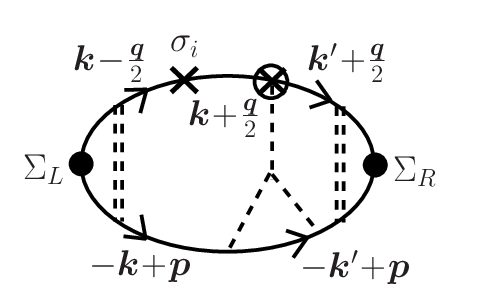}
 \caption{A diagram for the spin accumulation induced by the spin Hall effect induced by random spin-orbit interaction (denoted by $\otimes$) due to impurity in the diffusive case. Spin vertex is denoted by $\times$ and black circles denote self-energy due to the superconductor. Single dotted line denotes the impurity scattering, while the double dotted line is a ladder of impurity scattering, the cooperon, describing the diffusion of a spin-singlet electron pair.
  \label{FIGshSNSdif}}
\end{figure}

\begin{figure}\centering
\includegraphics[width=0.45\hsize]{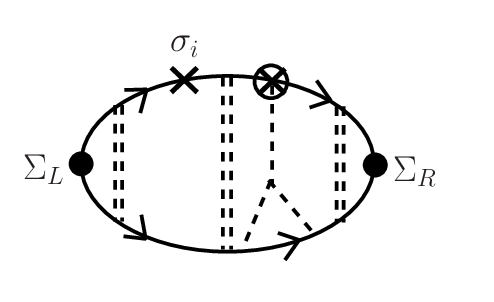}
 \raisebox{2\baselineskip}{\Large{$=0$}}
 \caption{A diagram with three cooperons separating the spin  and spin-orbit vertices vanishes.
  \label{FIGshSNSdif0}}
\end{figure}

%%%%%%%%%%%%%%%%%%%%%%%%%%%%%%%%%%%%%%%%%%%
In the model with random impurity, the dominant contribution at long wavelength comes from diagrams involving electron-pair propagation with successive impurity scattering,  the cooperon.
The contributions with cooperons are shown in Fig. \ref{FIGshSNSdif}.
The contribution with a cooperon inserted between the spin vertex and the spin-orbit  interaction vanishes (Fig. \ref{FIGshSNSdif0}).
The cooperon wave vectors for the two cooperons are $\pv+\frac{\qv}{2}$ and  $\pv-\frac{\qv}{2}$, as read from the diagram.
The response function $K_{s,i}$ is thus modified to be (using $2\pi\dos\tau\nimp\vimp^2=1$)
\begin{align}
\tilde{K}_{s,i}(\qv,\pv,\omega) &= K_{s,i}(\qv,\pv,\omega) \nnr &
\times \frac{1}{\tau^2}
\frac{1}{D\lt(p+\frac{\qv}{2}\rt)^2-2i\omega}
\frac{1}{D\lt(p-\frac{\qv}{2}\rt)^2-2i\omega}.
  \end{align}

%%%%%%%%%%%%%%%%
%%%%%%%%%%%%%%%%%%%%%%%%%%%%%%%%%%%%%
%%%%%%%%%%%%%%%%
%%%%%%%%%%%%%%%%%%%%%%%%%%%%%%%%%%%%%
The spin density induced by the spin Hall effect is
\begin{align}
 s_i(\qv)
& =
i\sumom (2f(\omega)-1) [\Sigma(\omega)]^2 \sum_{\pv}
(e^{i\varphi} e^{i\pv\cdot\Lv} + e^{-i\varphi}  e^{-i\pv\cdot\Lv}) \nnr & \times
 \frac{K_{s,i}(\qv,\pv,\omega)}{\tau^2}
\frac{1}{D\lt(p+\frac{\qv}{2}\rt)^2-2i\omega}
\frac{1}{D\lt(p-\frac{\qv}{2}\rt)^2-2i\omega}.
 \label{shDif1}
\end{align}
Defining
\begin{align}
 \Gamma_j(\qv,\omega)
& =
 \sum_{\pv}
 p_j e^{i\pv\cdot\Lv}
\frac{1}{D\lt(p+\frac{\qv}{2}\rt)^2-2i\omega}
\frac{1}{D\lt(p-\frac{\qv}{2}\rt)^2-2i\omega},
 \label{PiDif}
\end{align}
and noting that this function is odd in inversion along the $x$ axis, the spin density is
\begin{align}
 \sv(\rv)
 =&
10\pi^2 \lambdaso \dos^2 \vimp mD \sumom (2f(\omega)-1) [\Sigma(\omega)]^2
\nnr & \times
\lt(\nabla\times \bm{\Gamma}(\rv,\omega) \rt) \sin\varphi.
 \label{shDif2}
\end{align}
Using the result of the Josephson current, (\ref{JJDifq2}), we obtain
\begin{align}
\sv(\rv) &=  \lambda_{\rm sh}(\nabla \times \jv_{\rm J}).
\label{shjj}
\end{align}
where
\begin{align}
  \lambda_{\rm sh}
  = \frac{5}{4}\pi \lambdaso \dos\vimp m
  =\frac{5}{8}\pi \epsilon_{\rm so} \frac{\dos}{\ef},
\end{align}
is the spin Hall coefficient for the supercurrent, $\epsilon_{\rm so} \equiv  \lambdaso \vimp \kf^2$ being the energy of the spin-orbit interaction.
Thus, the supercurrent-induced spin Hall effect has the same form as in the normal case, Eq. (\ref{sh}), resulting in a spin accumulation that varies spatially perpendicular to the current and spin directions.

%%%%%%%%%%%%%%%%%%%%%%%%%%%%%%%%%%%%%%%%%%%%%
\section{The inverse spin Hall effect \label{SECish}}
%%%%%%%%%%%%%%%%%%%%%%%%%%%%%%%%%%%%%%%%%%%%%
The inverse spin Hall effect is studied by calculating the supercurrent when a spatially modulated magnetic field, $\Bv_{s}$, is applied.
The spatial gradient of the magnetic field is a source term for a spin current in the conventional picture.
We write the interaction Hamiltonian as
\begin{align}
H_{B}= \gamma \intr \Bv_{s}\cdot c^\dagger \sigmav c.
 \end{align}
We thus include an additional current vertex in the diagram for the spin Hall effect and evaluate the response to linear order in the external wave vector $\qv$.
As the current vertex is odd in the wave vector, the response function is even in the cooperon wave vector $\pv$, and is finite at $\pv=0$.
The inverse spin Hall current is thus proportional to $\cos\varphi$, resulting in a phase shift.

%%%%%%%%%%%%%%%%%%%%%%%%%%%%%%%%%%%%%%%%%%%%%
\subsection{Ballistic case}
\begin{figure}\centering
\includegraphics[width=0.6\hsize]{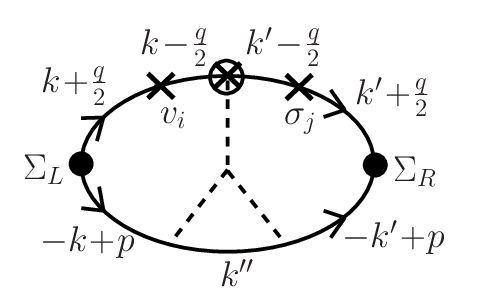}
 \caption{ Diagram contributing to the inverse spin Hall effect in the ballistic limit. A current vertex is denoted by $v_i$, $i=x,y,z$ being the direction.  An external wave vector $\qv$ is injected by the applied magnetic field and is absorbed at the current vertex.
 The diagram with $v_i$ and $\sigma_j$ interchanged yields the same result.
 A diagram with both $v_i$ and $\sigma_j$ on either side of the spin-orbit interaction does not contribute to the inverse spin Hall effect, as it is even in $\qv$.
  \label{FIGish_bal}}
\end{figure}

Diagram  contributing to the response in the ballistic case is shown in Fig. \ref{FIGish_bal}.
%%%%%%%%%%%%%%%%
%%%%%%%%%%%%%%%%%%%%%%%%%%%%%%%%%%%%%
For the current along the $i$ direction and $j$ being the direction of the applied magnetic field, the bare response function is
\begin{align}
 K_{{\rm ish}}^{ij}& (\qv,\pv,\omega) =
  i\lambdaso\nimp\vimp^3
 \sum_{\kv\kv'\kv''}
 \lt[\lt(\kv'+\frac{\qv}{2}\rt)\times\lt(\kv+\frac{\qv}{2}\rt)\rt]_j
 \nnr & \times
  g_{\kv'+\frac{\qv}{2}}^\adv(\omega)v_i g_{\kv'+\frac{\qv}{2}}^\adv(\omega)g_{\kv+\frac{\qv}{2}}^\adv(\omega)  g_{\kv-\frac{\qv}{2}}^\adv(\omega)
  \nnr & \times
  g_{-\kv'+\pv}^\ret (-\omega)g_{-\kv+\pv}^\ret (-\omega)g_{\kv''}^\ret (-\omega).
  \label{KishRSO}
  \end{align}
The contribution linear in $\qv$ for small $\pv$ is
\begin{align}
 K_{{\rm ish}}^{ij}& (\qv,\pv,\omega) =
  \pi \dos \lambdaso\nimp\vimp^3 \frac{q_k}{2m} \epsilon_{jkl}
 \sum_{\kv\kv'} k'_i(k-k')_l
 \nnr & \times
  (g_{\kv'}^\adv(\omega))^2 (g_{\kv}^\adv(\omega))^2
  g_{-\kv'+\pv}^\ret (-\omega)g_{-\kv+\pv}^\ret (-\omega)
  \nnr
   &=
   \lambdaso \pi^2 \dos^2 \vimp m D \tau^2  \epsilon_{ijk}q_k.
%   \lambdaso 2\pi \dos \kf^2 \tau^3 \frac{1}{6m} \epsilon_{ijk}q_k
\end{align}
We see here that the response function is even in $\pv$, meaning  it sees the two superconductors symmetrically, resulting in the $\cos\varphi$-dependence.

%%%%%%%%%%%%%%%%
\begin{figure}\centering
\includegraphics[width=0.6\hsize]{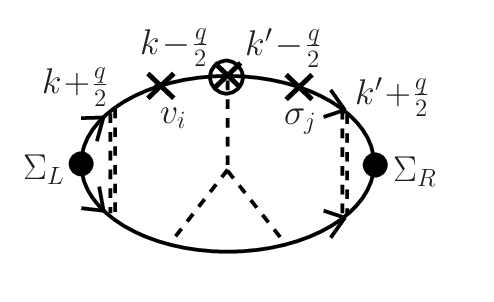}
 \caption{ Diagram contributing to the diffusive ISHE at the second order of the self-energy. A current vertex is denoted by $v_i$.  An external wave vector $\qv$ is injected by the applied magnetic field and is absorbed at the current vertex.
The response function $K_{\rm ish}^{ij}$ in Eq. (\ref{ishDif10}) represents the central part of the diagram without the cooperons.
  \label{FIGish}}
\end{figure}
%%%%%%%%%%%%%%%%%%%%%%%%%%%%%%%%%%%%%

%%%%%%%%%%%%%%%%%%%%%%%%%%%%555%%%%%
\subsection{Diffusive case}
%%%%%%%%%%%%%%%%%%%%%%%%%%%%%
Including two cooperons, the inverse spin Hall supercurrent  in the diffusive case (Fig. \ref{FIGish}) is
\begin{align}
 j_{\rm ish}^i(\qv)
& =
i B_{s,j}(\qv)\sumom (2f(\omega)-1) [\Sigma(\omega)]^2
\nnr &\times \sum_{\pv}
(e^{i\varphi} e^{i\pv\cdot\Lv} + e^{-i\varphi}  e^{-i\pv\cdot\Lv})
 \frac{K_{\rm ish}^{ij}(\qv,\pv,\omega)}{\tau^2}
 \nnr & \times
\frac{1}{D\lt(p+\frac{\qv}{2}\rt)^2-2i\omega}
\frac{1}{D\lt(p-\frac{\qv}{2}\rt)^2-2i\omega}.
 \label{ishDif10}
\end{align}
Defining the second-order cooperon propagator as
\begin{align}
 \Gamma_2(L,\omega)
& =
 \sum_{\pv}
 e^{i\pv\cdot\Lv}
\frac{1}{(Dp^2-2i\omega)^2},
 \label{Pi0Dif}
\end{align}
the inverse spin Hall current reads
\begin{align}
 \jv_{\rm ish}
= & \pi^2 \lambdaso \dos^2 \vimp m D
 (\nabla\times \Bv_{s})\cos\varphi
 \nnr &\times \sumom (2f(\omega)-1) [\Sigma(\omega)]^2
  \Gamma_2(L,\omega).
 \label{ishDif1}
\end{align}
The $\omega$-integration is carried out neglecting $\omega$-dependence of $\Sigma(\omega)$ as
\begin{align}
\sumom & (2f(\omega)-1)  \Gamma_2(L,\omega)
 \nnr &= \frac{i}{8\pi^2D^2} \int_0^\infty d\omega \frac{1}{\sqrt{\omega/D}} \tanh\frac{\beta \omega}{2} e^{-\sqrt{\omega/D}L}\nnr & \times (\cos \sqrt{\omega/D}L+\sin \sqrt{\omega/D}L)
 \nnr
 &=\frac{i}{4\pi^2DL}F(L/\ell_T),
\end{align}
where the function $F(\mu)$ is identical to the one governing the Josephson current, Eq. (\ref{Fmudef}).
The inverse spin Hall current is, therefore,
\begin{align}
 \jv_{\rm ish} &= \gamma \lambda_{\rm ish}(\nabla\times \Bv_{s})\cos\varphi,
\end{align}
where
\begin{align}
 \lambda_{\rm ish} &
   =   \frac{1}{8}\epsilon_{\rm so}\dos\frac{\dos \Sigma^2}{\ef L} F(L/\ell_T).
\end{align}

The total current, including the self-energy to second order is, therefore,
\begin{align}
 j &= j_{c2} \sin(\varphi+\delta\varphi),
 \label{jwithphaseshift2}
\end{align}
where
\begin{align}
\delta \varphi %= \frac{\lambda_{\rm ish}}{c}\nabla_y B_z
= \frac{ \lambda_{\rm ish} }{j_{c2}}\gamma \nabla_y B_{s,z}
= -\frac{ \pi }{16}\epsilon_{\rm so}\dos\frac{L}{\ef }\gamma \nabla_y B_{s,z},
\label{deltavarphi2}
\end{align}
is the phase shift.
Equations (\ref{jwithphaseshift2}) and (\ref{deltavarphi2}) constitute the central result of this paper.
Superconductors thus absorb the supercurrent induced by the inverse spin Hall effect by adjusting their phase, in the same way as  the inverse Rashba-Edelstein effect \cite{Konschelle15}.
There is, however, a significant feature of the ISHE: the phase shift does not require broken structural inversion symmetry. Instead, the gradient of the applied magnetic field, which generates an inhomogeneous spin accumulation equivalent to an  equilibrium spin current, is sufficient.

%%%%%%%%%%%%%%%%%%%%%%%%%%%%%%%%%%%%%%%%%%%%%%%%%%%%%%%%%%%%%%
\section{Higher harmonics of the inverse spin Hall effect \label{SEChigher}}
\begin{figure}\centering
\includegraphics[width=0.6\hsize]{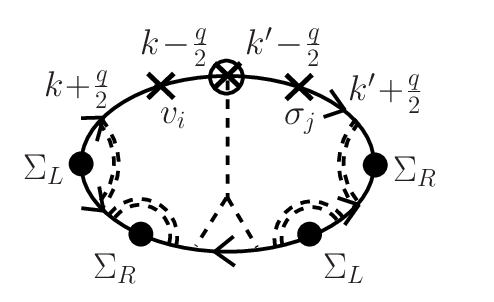}
 \caption{ Fourth-order diagram contributing to the inverse spin Hall effect in the diffusive case.
  \label{FIGishdif4}}
\end{figure}

\begin{figure}\centering
\includegraphics[width=0.4\hsize]{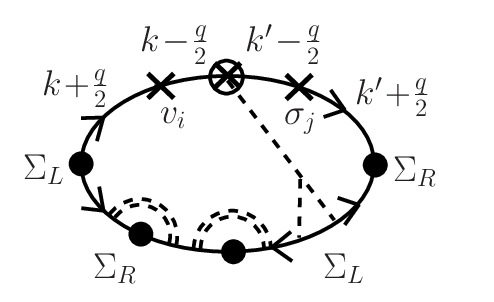}
 \raisebox{2\baselineskip}{\Large{$+$}}
\includegraphics[width=0.4\hsize]{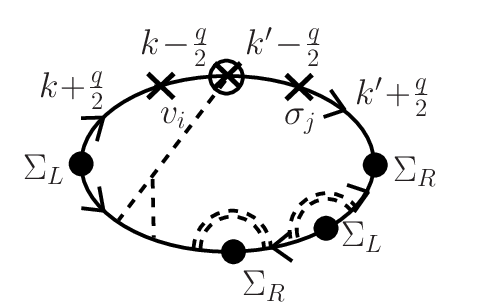}
 \raisebox{2\baselineskip}{\Large{$=0$}}
 \caption{ Fourth-order diagrams that cancel each other.
  \label{FIGishdif40}}
\end{figure}

%%%%%%%%%%%%%%%%%%%%%%%%%%%%%%%%%%%%%%%%%%%%%%%%%%%%%%%%%%%%%%
The anomalous phase shift, which indicates a directional current, might naively be expected to lead to a Josephson diode effect, in which the critical current depends on the direction of the current. This is not, however, the case.
%, since a static anomalous phase shift is a ground-state property and does not affect the response to an injected current.
A diode effect emerges when higher harmonics arising from multiple interface scattering are taken into account \cite{Reinhardt24}.
In our scheme, the fourth-order contribution to the ISHE is accounted for by including the process shown in Fig. \ref{FIGishdif4}.
Other processes, as shown in Fig. \ref{FIGishdif40}, vanish.
%%%%%%%%%%%%%%%%
%%%%%%%%%%%%%%%%%%%%%%%%%%%%%%%%%%%%%
%%%%%%%%%%%%%%%%
%%%%%%%%%%%%%%%%%%%%%%%%%%%%%%%%%%%%%

The response function of the inverse spin Hall current, including the self-energy to fourth order, is with two extra Green's functions compared to $ K_{{\rm ish}}^{ij}$ as
%(removing extra self-energies $\Sigma^2$ and two cooperons modifying them)
\begin{align}
 K_{{\rm ish}}^{(4)ij}&(\qv,\pv,\omega) =
  i\lambdaso\nimp\vimp^3
 \sum_{\kv\kv'\kv''}
 \lt[\lt(\kv'+\frac{\qv}{2}\rt)\times\lt(\kv+\frac{\qv}{2}\rt)\rt]_j
 \nnr & \times
  g_{\kv'+\frac{\qv}{2}}^\adv(\omega)v_i g_{\kv'+\frac{\qv}{2}}^\adv(\omega)g_{\kv+\frac{\qv}{2}}^\adv(\omega)  g_{\kv-\frac{\qv}{2}}^\adv(\omega) \nnr &\times
  g_{-\kv'+\pv}^\ret (-\omega)g_{-\kv+\pv}^\ret (-\omega)
  g_{\kv'-\pv+\pv_1}^\adv
  g_{\kv-\pv+\pv_1}^\adv
  g_{\kv''}^\ret (-\omega),
  \end{align}
which is calculated to obtain
$ K_{{\rm ish}}^{(4)ij} \simeq -\tau^2  K_{{\rm ish}}^{ij}$.
The fourth-order inverse spin Hall current is therefore
\begin{align}
j_{\rm ish}^{(4)} &= \gamma \lambda_{\rm ish}^{(4)}( \nabla_y B_z)\cos 2\varphi,
\end{align}
with
\begin{align}
 \lambda_{\rm ish}^{(4)} &= \lambda_{\rm ish}
 \frac{1}{2^6}\lt(\frac{\Sigma L^2}{D} \rt)^2 \frac{\tilde{F_4}(L/\ell_T)}{F(L/\ell_T)}.
\end{align}
Here the factor
\begin{align}
  \tilde{F_4} (\mu) & \equiv \int_0^\infty dx \frac{e^{-x}}{x^4}\tanh \lt(\frac{x^2}{2\mu^2}\rt)
  \nnr & \times
  \lt(\lt(1+2x+\frac{2}{3}x^2\rt)\cos x-\lt(1-\frac{2}{3}x^2\rt)\sin x\rt),
\end{align}
arose from the $\omega$-integration of the  four cooperon amplitudes,
\begin{align}
 \Gamma_4 & \equiv \sum_{\pv}\frac{e^{i\pv\cdot\Lv}}{(Dp^2-2i\omega)^4}
  \nnr
  &= -i\frac{1\pm i}{2^9 \pi D^4} \lt(\frac{D}{|\omega|}\rt)^{5/2}
  \nnr & \times
  \lt(1+ L\sqrt{\frac{|\omega|}{D}} \mp i L\sqrt{\frac{|\omega|}{D}} \lt(1+\frac{2}{3}L\sqrt{\frac{|\omega|}{D}} \rt)\rt),
\end{align}
as
\begin{align}
 \sumom (2f_\omega-1) \Gamma_{4} &=
   - i\frac{L^3}{2^8\pi^2 D^3 } \tilde{F_4}(L/\ell_T).
\end{align}

The total supercurrent to fourth-order,
$j^{(2,4)}\equiv j_{\rm J}+j_{\rm J}^{(4)}+j_{\rm ish}+j_{\rm ish}^{(4)}$
is, therefore,
\begin{align}
 j^{(2,4)}
=j_{c2} \sin (\varphi+\delta\varphi)+j_{c4} \sin (2\varphi+\delta\varphi^{(4)}),
\label{totaljdif}
\end{align}
with $\delta\varphi$ of (\ref{deltavarphi2}), and the fourth-order phase shift is
\begin{align}
 \delta\varphi^{(4)} &= \frac{\lambda_{\rm ish}^{(4)}}{j_{c4}}\nabla_y B_{s,z},
 \label{phi4}
\end{align}
where
\begin{align}
  \frac{\lambda_{\rm ish}^{(4)}}{j_{c4}} &= -\frac{1}{4} \frac{\tilde{F_4}}{F}  \frac{\lambda_{\rm ish}}{j_{c2}}.
  \label{24ratio}
\end{align}

%%%%%%%%%%%%%%%%%%%%%%%%%%%%%%%%%%%%%%%%%%%%%%
%\section{Diode effect}
\section{Inverse spin Hall-induced Josephson diode effect \label{SECdiode}}
\begin{figure}\centering
\includegraphics[width=0.9\hsize]{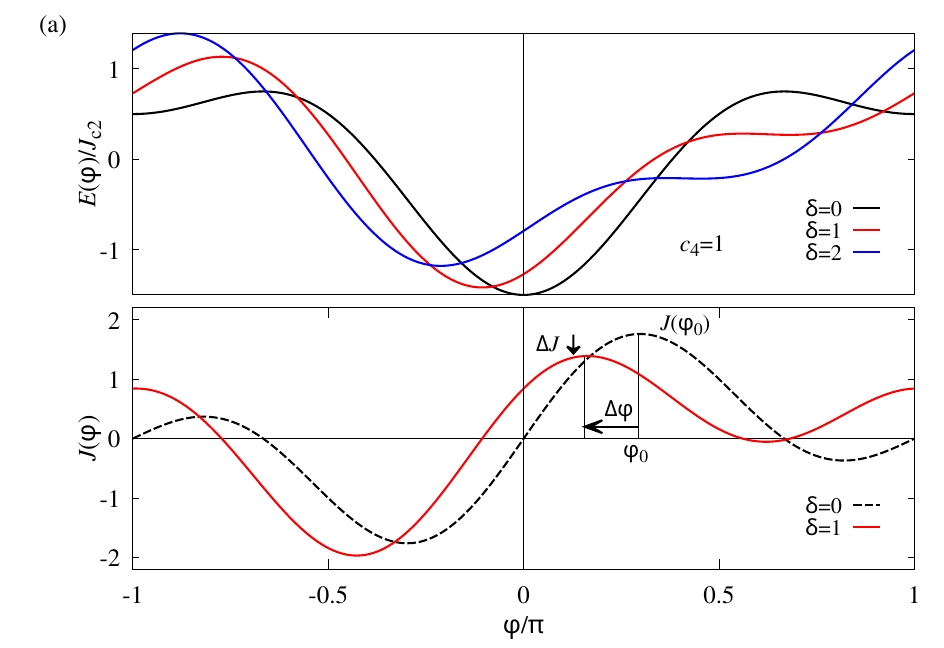}
\includegraphics[width=0.6\hsize]{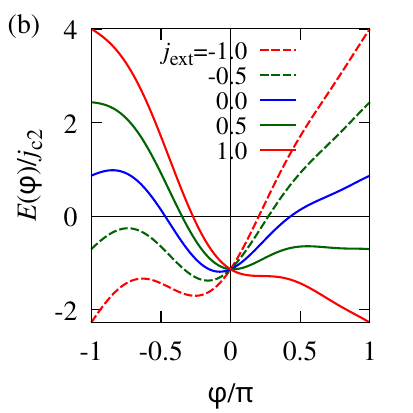}
 \caption{ (a): Upper panel: Potential $E(\varphi)$ including the second harmonic with $c_4=1$ normalized by $j_{c2}$ as a function of the phase difference $\varphi$ with and without phase shift $\delta$. Finite $\delta$ induces an asymmetry of the potential.  For clarity, $c_4$ and $\delta$ are chosen to be relatively large.
 Lower panel: The function $J(\varphi)$ representing the Josephson current. The maximum of $J$ for finite $\delta$ is at $\varphi=\varphi_0+\Delta\varphi$, where $\varphi_0$ corresponds to the maximum at $\delta=0$. The maximum value $J(\varphi_0+\Delta\varphi)$ differs for positive and negative $\varphi$, giving rise to the Josephson diode effect.
 (b): Tilted potential with $c_4=0.5$ and $\delta=1$ under an external current $j_{\rm ext}$ (normalized by $j_{c2}$).
Due to the asymmetry, the critical current, the current at which  the barrier vanishes, differs for positive and negative currents. We see that $j_{\rm ext}=1.0$ is above the positive critical current, while
$j_{\rm ext}=-1.0$ is still in the superconducting regime with a barrier.
  \label{FIGpot}}
\end{figure}

%%%%%%%%%%%%%%%%%%%%%%%%%%%%%%%%%%%%%%%%%%%%%%
The phase potential $E(\varphi)$ is defined so that its derivative is
the supercurrent;
$j^{(2,4)}=\frac{d}{d\varphi}E(\varphi)$.
The potential for the total current (\ref{totaljdif}) is (with a shift of the origin of $\varphi$)
\begin{align}
 E(\varphi) &= j_{c2}\lt( - \cos \varphi - \frac{{c}_4}{2}\cos (2\varphi+\delta) \rt),
 \label{E24}
\end{align}
where  ${c}_4=j_{c4}/j_{c2}$, and $\delta=\delta\varphi^{(4)}-2\delta\varphi$.
Our results, Eqs. (\ref{phi4})(\ref{24ratio}), indicates $\delta\varphi^{(4)}\neq 2\delta\varphi$, meaning that the inverse spin Hall effect does induce an asymmetry of the potential (Fig. \ref{FIGpot}(a)).

%%%%%%%%%%%%%%%%
%%%%%%%%%%%%%%%%%%%%%%%%%%%%%%%%%%%%%

When an electric current $j_{\rm ext}$ is applied to the junction, the potential is tilted, to be (Fig. \ref{FIGpot}(b))
\begin{align}
 E(\varphi) &=
   j_{c2}\lt( - \cos \varphi - \frac{{c}_4}{2}\cos (2\varphi+\delta) \rt)-j_{\rm ext}\varphi.
 \label{E24}
\end{align}
When the external current exceeds the maximum of the intrinsic current, superconductivity vanishes.
The critical current, $j^{\rm cr}$, is thus given by the maximum of $ j_{c2}|J(\varphi)|$, where
\begin{align}
 J(\varphi) &=
    \sin \varphi + {c}_4\sin (2\varphi+\delta).
 \label{Jdef}
\end{align}
Due to the asymmetry introduced by $\delta$ and $c_4$, the critical current differs in the positive and negative directions, resulting in the Josephson  diode effect.

Let us study the case of small $\delta$.
In the absence of $\delta$, the maximum current for positive and negative directions are  at $\varphi=\pm \varphi_0$, where $\varphi_0$ is determined by $J'(\varphi_0)=0$, i.e.,
\begin{align}
 \cos(\varphi_0) = -2{c}_4 \cos (2\varphi_0).
\end{align}
The critical current is thus $j^{{\rm cr},\pm}|_{\delta=0}/j_{c2}=\pm |J(\varphi_0)|$ for positive and negative currents.
Including the phase shift to the linear order, the minimum is modified to be at $\pm\varphi_0+\Delta\varphi$
(defined by $J'(\pm\varphi_0+\Delta\varphi)=0$), where
\begin{align}
 \Delta\varphi &= -\frac{4 c_4 \cos\varphi_0}{1+8 c_4 \cos\varphi_0}\delta.
\end{align}
Accordingly, the maximum of the function $|J(\varphi)|$ is modified  to be
\begin{align}
 J(\pm\varphi_0+\Delta\varphi) &=  J(\pm\varphi_0) +\Delta J,
\end{align}
where the shift is
\begin{align}
 \Delta J
 &=
 c_4 \delta \cos 2\varphi_0
\end{align}
The magnitude of the critical current is
$|j^{{\rm cr},\pm}|/j_{c2}=|J(\varphi_0)|\pm \Delta J$, and the difference of the positive and negative critical currents,
$\Delta j^{{\rm cr}}\equiv j^{{\rm cr},+}-j^{{\rm cr},-}$, is
\begin{align}
 \frac{\Delta j^{{\rm cr}}}{j_{c2}} &= 2\Delta J
  = -\delta \frac{\sqrt{1+2^5 (c_4)^2}-1}{8c_4}.
\end{align}

The ISH diode effect arises in general systems even without structural  symmetry breaking.
The spin current, represented by an inhomogeneous magnetic field, is sufficient to break  inversion and time-reversal symmetries and induces the diode effect.

\subsection{Field gradient realized by a magnetic dot}
%%%%%%%%%%%%%%%%
\begin{figure}\centering
\includegraphics[width=0.45\hsize]{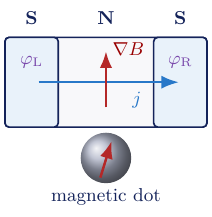}
 \caption{Schematic picture showing a magnetic  field gradient induced by attaching a magnetic dot to a side of the SNS junction.
  \label{FIGSNSdot}}
\end{figure}
%%%%%%%%%%%%%%%%%%%%%%%%%%%%%%%%%%%%%
A practical way to apply a field gradient is to put a magnetic dot near one of the edges of N (Fig. \ref{FIGSNSdot}).
If the spin polarization of the electron in N near the dot is induced by an exchange interaction $J_{\rm ex}$, and if the polarization decays on the length scale of the spin diffusion length, we have $\gamma \nabla_y B_{\rm s}\sim J_{\rm ex}/\ell_{\rm s}$.
The phase shift, Eq. (\ref{deltavarphi2}) reads in this case
$|\delta\varphi| =
\frac{\pi}{16}\frac{J_{\rm ex}}{\ef}\frac{\epsilon_{\rm so}}{\ef}\frac{L}{\ell_{\rm s}}$.
For $L/\ell_{\rm s}\sim 1$, $\frac{J_{\rm ex}}{\ef}\sim 1$ and $\frac{\epsilon_{\rm so}}{\ef}\sim 1$
%{\color{blue}
($\frac{\epsilon_{\rm so}}{\ef}\sim \theta_{\rm sh}$ in terms of the spin Hall angle (Eq. (\ref{spinhallangle}))),
%},
we have $|\delta\varphi| \sim 0.2$.
The parameter determining higher harmonics is estimated as
$c_4\sim \frac{9}{16}(\Sigma\tau)^2\lt(\frac{L}{\ell}\rt)^4$, where $\ell$ is the mean free path.
The self-energy is $\sigma\sim\ef$ if the hopping amplitude at the SN interface is $t\sim\ef$.
If $L/\ell=3$, $t/\ef=0.1$ and $\ef\tau=10$, we have $c_4\sim 0.5$, resulting in
$ \delta j_c /j_{c2} \sim 0.1$ for  $|\delta\varphi| \sim 0.2$.
This nonreciprocality would be sufficiently large for experimental detection.

%%%%%%%%%%%%%%%%%%%%%%%%%%%%%%%%%%%%%%%%%%%%%%%%%%%%%%%%%%%%%%
%%%%%%%%%%%%%%%%%%%%%%%%%%%%%%%%%%%%%%%%%%%%%%%%%%%%%%%%%%%%%%
\section{Discussion \label{SECdiscussion}}
%%%%%%%%%%%%%%%%%%%%%%%%%%%%%%%%%%%%%%%%%%%%%%%%%%%%%%%%%%%%%%

Let us look into the result of supercurrent-induced spin Hall effect in the context of Sec. \ref{SECreciprocal}.
Mapping an SNS with a small phase difference $\varphi$ to a continuum system, $\sin\varphi$ is replaced by  $L \Av_\varphi$, where
$\Av_\varphi\equiv\nabla\varphi$ is a vector potential.
The Josephson current is proportional to the vector potential as
\begin{align}
\jv=\tilde{j}_{c2} \Av_\varphi,
\end{align}
 where $\tilde{j}_{c2}\equiv {j_{c2}}L$.
The SH response (\ref{shjj}) reads ($\tilde{\lambda}_{\rm sh}\equiv {\lambda_{\rm sh}}L$)
\begin{align}
\sv=\tilde{\lambda}_{\rm sh}\nabla\times\Av_\varphi,
\end{align}
The spin Hall effect induced by a supercurrent is, therefore,  induced by a magnetic field $\Bv_{\varphi}\equiv \nabla\times\Av_\varphi$ generated by the superconducting phase,  and it corresponds to the first static term of Eq. (\ref{sh}).

The inverse spin Hall effect, Eq. (\ref{jwithphaseshift2}), indicates that an anomalous vector potential
\begin{align}
 \delta \Av_\varphi\equiv \delta\varphi/L=\tilde{\lambda}_{\rm ish}(\nabla\times\Bv_{s}),
\end{align}
where
$\tilde{\lambda}_{\rm ish}\equiv \gamma\frac{ \lambda_{\rm ish} }{j_{c2}L} $
is induced.

The suprecurrent-induced direct and inverse spin Hall effects, therefore, correspond to static responses represented by the term $\lambda_0$ of Eqs.
(\ref{sh})(\ref{ish}), while those in dissipative normal metals are represented by the dynamics  term $\lambda_1$.

%%%%%%%%%%%%%%%%%%%%%%%%%%%%%%%%%%%%%%%%%%%%%%%%%%%%%%%%%%%%%%
%%%%%%%%%%%%%%%%%%%%%%%%%%%%%%%%%%%%%%%%%%%%%%%%%%%%%%%%%%%
\section{Rashba-Edelstein effect \label{SECrashba}}
%%%%%%%%%%%%%%%%%%%%%%%%%%%%%%%%%%%%%%%%%%%%%%%%%%%%%%%%%%%
%
For the sake of comparison, in the case of the Rashba spin-orbit interaction, a phase shift is induced  by the field along the $y$ direction due to the inverse Rashba-Edelstein effect \cite{Konschelle15}, in addition to the ISHE.
We consider the Rashba vector along the $z$ axis, i.e., perpendicular to the junction.
The Hamiltonian for N is spin-dependent as
\begin{align}
  H_{\rm N}&=\sum_{\kv} c^\dagger_{\kv} \lt(\ekv+\gammav_\kv\cdot\sigmav\rt) c_{\kv},
\end{align}
where
\begin{align}
\gammav_\kv= \alpha(\hat{\zv}\times\kv),
\end{align}
$\alpha$ being the Rashba coefficient.
The velocity operator is
\begin{align}
\hat{\vv}=\frac{\kv}{m}- \alpha(\hat{\zv}\times\sigmav).
\end{align}
We consider the ballistic case.
The retarded component of the Rashba Green's function in the clean limit is
\begin{align}
 G^\ret_{\kv,\omega} &=
  \frac{1}{\omega-\epsilon_{\kv}-\gammav_\kv\cdot\sigmav+i0},
\end{align}
where $i0$ is an infinitesimal positive imaginary part.
It is useful to introduce spin-diagonal Green's functions and write
\begin{align}
 G^\ret_{\kv,\omega} &= f^\ret_{\kv\omega}-\frac{{\gammav}_\kv}{|\gammav_\kv|}\cdot\sigmav h^\ret_{\kv\omega},
\end{align}
where
\begin{align}
f^\ret_{\kv\omega} &=\frac{1}{2} \sum_{\sigma=\pm} g_{\kv\sigma\omega}^{{\rm (R)}\ret} , &
h^\ret_{\kv\omega} &=\frac{1}{2} \sum_{\sigma} \sigma g_{\kv\sigma\omega}^{{\rm (R)}\ret}.
\label{fhdef}
\end{align}
Here
\begin{align}
g_{\kv\sigma\omega}^{{\rm (R)}\ret}
&=
  \frac{1}{\omega-\epsilon_{\kv\sigma}+i0},
\end{align}
is a diagonalized Rashba Green's function, with $\epsilon_{\kv\sigma}\equiv\epsilon_{\kv} -\sigma|\gammav_\kv|$ ($\sigma=\pm 1$).

For the present spin-dependent Green's function, the spin-singlet nature of the superconductor needs to be taken into account.
Writing  the spin suffix  at interfaces $\sigma_R,\sigma_L=\pm$ explicitly, a static expectation value of an observable $\hat{O}$ with finite wave vector $\qv$ is written as
\begin{align}
 O_{\qv}
& =
i\sumom (2f(\omega)-1) \Sigma(\omega)^2
\sum_{\sigma_R,\sigma_L}\sum_{\kv\pv}
\nnr
 &\times
 \biggl[
 e^{i\varphi} \lt[G^\adv(\kv+\frac{\qv}{2},\omega) \hat{O} G^\adv(\kv-\frac{\qv}{2},\omega)  \rt]_{\sigma_R,\sigma_L}
 \nnr &\times
  G^\ret_{-\sigma_R,-\sigma_L}(-\kv+\pv,-\omega)
+({\rm L}\leftrightarrow{\rm R})
 \biggr].
 \label{GNLR2}
\end{align}
The spin summation is written by use of the Pauli matrix as (for any $2\times2$ matrices $A$ and $B$)
\begin{align}
\sum_{\sigma\sigma'}A_{\sigma\sigma'}B_{-\sigma,-\sigma'}
&= \tr[\sigma_x A \sigma_x B^{\rm t}],
    \label{singlettrace}
\end{align}
where $^{\rm t}$ denotes transpose.
We thus have
\begin{align}
 O_{\qv}
& =
i\sumom (2f(\omega)-1) \Sigma(\omega)^2
\nnr
 &\times \sum_{\kv\pv}e^{i\pv\cdot\Lv}
 \tr \biggl[
 e^{i\varphi} G^\adv(\kv+\frac{\qv}{2},\omega) \hat{O} G^\adv(\kv-\frac{\qv}{2},\omega)
 \nnr & \times
  \sigma_x [G^\ret(-\kv+\pv,-\omega)]^{\rm t} \sigma_x
+({\rm L}\leftrightarrow{\rm R})
 \biggr].
 \label{Odef}
\end{align}
Luckily, however, in the case the Rashba vector is along the $z$ axis, we have
$ \sigma_x[G_{\kv,\omega}^\ret ]^{\rm t}\sigma_x=G_{\kv,\omega}^\ret$, and the response function is simplified.
The response function for the spin accumulation in Eq. (\ref{spin1}) is, therefore,
\begin{align}
K_{s,i}(\qv,\pv,\omega)
 & = \sum_{\kv}
\tr\biggl[
  G_{\kv+\frac{\qv}{2},\omega}^\adv \sigma_i G_{\kv-\frac{\qv}{2},\omega}^\adv
  G_{-\kv+\pv,-\omega}^\ret
\biggr].
  \label{Kspdef2}
\end{align}

Let us  first study the uniform spin polarization, $\qv=0$, corresponding to the Rashba-Edelstein effect.
At the linear order in the Rashba interaction, the result of the response function in the real space, $K_{s,y}(0,L,\omega)\equiv \sum_{\pv} e^{-i(\kv+\kv')\cdot \Lv}
K_{s,i}(0,\pv,\omega)$, is
\begin{align}
K_{s,y}(0,L,\omega)
&=
 4i\alpha m L\sum_{\kv\kv'}e^{-i(\kv+\kv')\cdot \Lv}
(g_{\kv\omega}^{\adv})^2 g_{\kv',-\omega}^{\ret},
\end{align}
where $g_{\kv\omega}^{\adv}$ denotes the Green's function neglecting the Rashba interaction,
and the $x$ and $z$ components vanish.
The spin density to the lowest order of the Rashba interaction  is thus
\begin{align}
 s_y &= \chi_{\rm R} j_{\rm J},
\end{align}
with
\begin{align}
 \chi_{\rm R} &= \frac{4m\alpha}{\ef},
\end{align}
being the Rashba-Edelstein coefficient for the supercurrent.

The inverse Rashba-Edelstein effect is similarly calculated.
The uniform ($q=0$) component of the response function is
\begin{align}
K_{\rm IRE}^{ij} (0,\pv,\omega) & =
\sum_{\kv}
 \tr \biggl[
 \nnr &
 (G^\adv_{\kv,\omega}\hat{v}_i  G^\adv_{\kv,\omega} \sigma_j  G^\adv_{\kv,\omega}
 + G^\adv_{\kv,\omega}\sigma_j G^\adv_{\kv,\omega}\hat{v}_i  G^\adv_{\kv,\omega})
 {G^\ret_{-\kv+\pv,-\omega}}
 \nnr
  &+ G^\adv_{\kv,\omega}\hat{v}_i  G^\adv_{\kv,\omega}
 G^\ret_{-\kv+\pv,-\omega}\sigma_j G^\ret_{-\kv+\pv,-\omega}
  \biggr].
\end{align}
Using $G^\adv_{\kv,\omega}\hat{v}_i  G^\adv_{\kv,\omega}=\partial_{k_i}  G^\adv_{\kv,\omega}$ and the integration by parts, we obtain
\begin{align}
K_{\rm IRE}^{ij} (0,\pv,\omega) & =
 -iL \delta_{ix} \sum_{\kv}
 \tr \biggl[
 G^\adv_{\kv,\omega} \sigma_j  G^\adv_{\kv,\omega} {G^\ret_{-\kv+\pv,-\omega}}
 \nnr &
  + G^\adv_{\kv,\omega}
 G^\ret_{-\kv+\pv,-\omega}\sigma_j G^\ret_{-\kv+\pv,-\omega}
  \biggr].
\end{align}
The inverse Rashba-Edelstein current to the linear order of the Rashba interaction turns out to be
\begin{align}
j_{{\rm IRE}}& = \chi_{\rm IRE} j_{c2}\gamma B_y \cos\varphi,
\end{align}
where
\begin{align}
 \chi_{\rm IRE} &= L\chi_{\rm RE}.
\end{align}

%%%%%%%%%%%%%%%%%%%%%%%%%%%%%%%%%%%%%%%%%%%%%%%%
\subsection{Spin Hall effect in Rashba system}
%%%%%%%%%%%%%%%%%%%%%%%%%%%%%%%%%%%%%%%%%%%%%%%%
The spin Hall effect due to the Rashba interaction can be analyzed in a similar manner. As pointed out in Ref. \cite{Shitade22}, the spin Hall effect in Rashba systems is subtle when formulated in terms of the spin current. In contrast, the present formulation based on the spin density provides a clear description.

To include a spatial variation of the induced spin density, we expand the response function (\ref{Kspdef2}) to linear order in the external wave vector $\qv$, to obtain (a superscript $(1)$ denotes the linear order in $\qv$)
\begin{align}
K^{(1)}_{s,i}(\qv,\pv,\omega)
 & = \frac{q_k}{2}\sum_{\kv}
\tr\biggl[
  G_{\kv,\omega}^\adv M_{ki} G_{\kv,\omega}^\adv  G_{-\kv+\pv,-\omega}^\ret
  \biggr],
  \label{Kshpdef2}
\end{align}
where
\begin{align}
M_{ki} & \equiv
\hat{v}_k G_{\kv,\omega}^\adv  \sigma_i - \sigma_i G_{\kv,\omega}^\adv \hat{v}_k.
\end{align}
It turns out that the linear contribution of the Rashba interaction vanishes, and the spin accumulation starts from the second order.
The response function is
\begin{align}
K^{(1)}_{s,i} & (\qv,\pv,\omega)
  = 4i\alpha^2\delta_{iz}q_k\epsilon_{klz}
 \sum_{\kv}
 \nnr &
 \lt[
  -k_l\lt((g_{\kv,\omega}^{\adv})^4+\frac{4k^2}{3m}(g_{\kv,\omega}^{\adv})^5\rt) g_{-\kv+\pv,-\omega}^{\ret}
  \rt. \nnr &  \lt.
  +\lt((g_{\kv,\omega}^{\adv})^3+\frac{2k^2}{3m}(g_{\kv,\omega}^{\adv})^4\rt) (-k+p)_l(g_{-\kv+\pv,-\omega}^{\ret})^2  \rt].
  \label{Kshres}
\end{align}
The result of the Rashba-induced spin Hall effect is
\begin{align}
 \sv_{\rm sh} &= \lambda_{\rm sh}^{\rm R}(\nabla\times \jv_{\rm J}),
\end{align}
where the spin Hall coefficient is
\begin{align}
\lambda_{\rm sh}^{\rm R} &=
   \frac{\alpha^2 m}{\ef^2}.
\end{align}

The inverse spin Hall effect due to the Rashba interaction  is studied by including an external wave vector $\qv$ representing the spatial variation in the response function for the inverse Rashba-Edelstein effect, $K_{\rm IRE}^{ij}$.
\begin{align}
K_{\rm ish}^{ij}&  (\qv,\pv,\omega) =
\sum_{\kv}
 \tr \biggl[
 g^\adv_{\kv+\frac{\qv}{2},\omega}\hat{v}_i  g^\adv_{\kv-\frac{\qv}{2},\omega}
 g^\ret_{\kv'+\frac{\qv}{2},-\omega}\sigma_j g^\ret_{\kv'-\frac{\qv}{2},-\omega}
\nnr& +
(g^\adv_{\kv+\frac{\qv}{2},\omega} \hat{v}_i  g^\adv_{\kv-\frac{\qv}{2},\omega} \sigma_j  g^\adv_{\kv+\frac{\qv}{2},\omega}
 +  g^\adv_{\kv-\frac{\qv}{2},\omega}\sigma_j  g^\adv_{\kv+\frac{\qv}{2},\omega}\hat{v}_i g^\adv_{\kv-\frac{\qv}{2},\omega}\hat{v}_k )
 \nnr & \times
 {g^\ret_{\kv',-\omega}}
 %\nnr  &
  \biggr]_{\kv'=-\kv+\pv}.
\end{align}
Expanding with respect to $\qv$, the  inverse spin Hall contribution (linear in $\qv$) turns out to be
\begin{align}
K_{\rm ish}^{(1)ij} (\qv,L,\omega) =&
-iL\delta_{ix}\frac{q_k}{2}\sum_{\kv\kv'}e^{i(\kv+\kv')\cdot\Lv}
\nnr &
 \tr \biggl[
G^\adv_{\kv,\omega} M_{kj}  G^\adv_{\kv,\omega}  {G^\ret_{\kv',-\omega}}
  \biggr]
  \nnr
  =& -iL\delta_{ix}K_{s,j}^{(1)} (\qv,L,\omega).
\end{align}
The inverse spin Hall  current is, therefore,
\begin{align}
 j_{\rm ish} &=
  \lambda^{\rm R}_{\rm ish} j_{c2} \gamma \nabla_y B_{s,z} \cos\varphi,
\end{align}
where $\lambda^{\rm R}_{\rm ish}=L\lambda_{\rm sh}^{\rm R}$.

Therefore, for the ballistic Rashba system, the second-order supercurrent  is
\begin{align}
 j= \gamma j_{c2} \cos\varphi (\chi_{\rm IRE}  B_{s,y}+\lambda^{\rm R}_{\rm ish}\nabla_y B_{s,z}).
\end{align}
The higher harmonics also exist for the Rashba case, resulting in a diode effect \cite{Reinhardt24}.
The phase in Rashba SNS can, therefore, be manipulated in two ways: by using the field along $y$ and $z$ directions.
The inverse Rashba-Edelstein effect reacts to the uniform component of the applied field, while the ISHE reacts to the spatial asymmetry of the field.

%%%%%%%%%%%%%%%%
\begin{figure}\centering
\includegraphics[width=0.8\hsize]{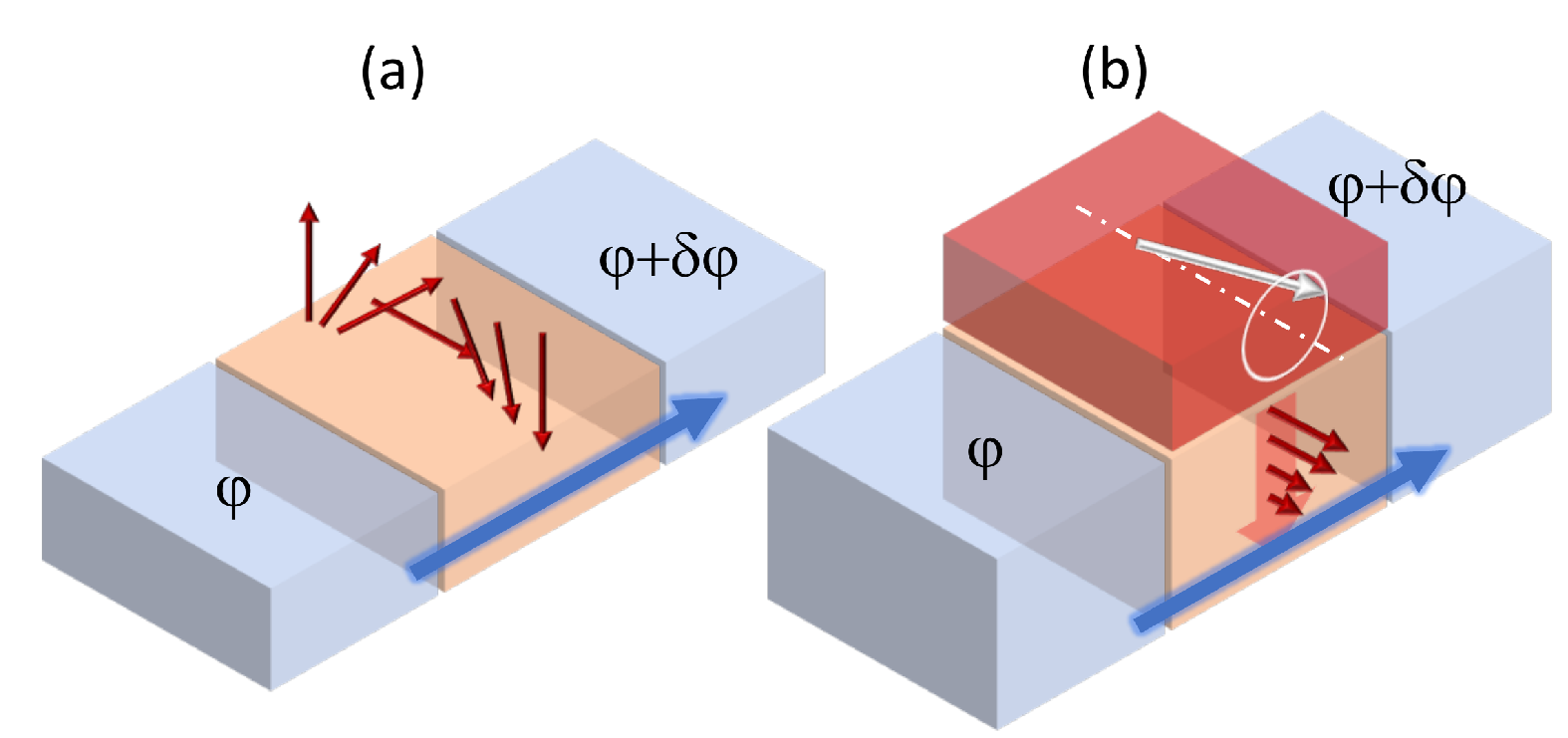}
 \caption{ Schematic figures showing the magnetization gradient generated by (a) a domain wall and (b) the spin pumping effect.
  \label{FIGdwsp}}
\end{figure}
%%%%%%%%%%%%%%%%%%%%%%%%%%%%%%%%%%%%%

\section{Summary \label{SECsummary}}
We have demonstrated that the ISHE in SNS Josephson junctions provides a powerful mechanism for phase control  \cite{MatsuoSA23}  via equilibrium spin currents generated by proximate magnetic textures. Such equilibrium spin currents may arise, for instance, from a proximate magnetic dot (Fig. \ref{FIGSNSdot}), a ferromagnet hosting a magnetic domain wall (Fig. \ref{FIGdwsp}(a)), or from an applied magnetic field gradient.
For a dot with quantum spin, the present SNS + magnetic dot systems work as  coupler between the Josephson phase Qbit and spin Qbit.
Beyond equilibrium scenarios, one can further envision the injection of nonequilibrium spin currents, for example, through spin pumping  (Fig. \ref{FIGdwsp}(b)).
Altogether, these possibilities open promising pathways for the magnetic control of Josephson devices, and offer  novel routes for detecting both equilibrium and nonequilibrium spin currents. These prospects motivate dedicated experimental efforts.

%%%%%%%%%%%%%%%%%%%%%%%%%%%%%%%%%%%%%%%%%%%%%%%%%%%%%%%%%%%
\acknowledgements
G. T. thanks for support by JSPS KAKENHI (Grant-in-Aid for Scientific Research (B) 26K00658) and
H. Kohno for useful discussion.
A. M. acknowledges support from the  France 2030 government investment plan managed by the French National Research Agency under grant reference PEPR SPIN – [SPINTHEORY] ANR-22-EXSP-0009 and [SUPERSPIN] ANR-24-EXSP-0012.
%%%%%%%%%%%%%%
\bibliography{/home/gt/References/24,/home/gt/References/gt,/home/gt/References/remarks,/home/gt/References/paper}

\begin{thebibliography}{38}%
\makeatletter
\providecommand \@ifxundefined [1]{%
 \@ifx{#1\undefined}
}%
\providecommand \@ifnum [1]{%
 \ifnum #1\expandafter \@firstoftwo
 \else \expandafter \@secondoftwo
 \fi
}%
\providecommand \@ifx [1]{%
 \ifx #1\expandafter \@firstoftwo
 \else \expandafter \@secondoftwo
 \fi
}%
\providecommand \natexlab [1]{#1}%
\providecommand \enquote  [1]{``#1''}%
\providecommand \bibnamefont  [1]{#1}%
\providecommand \bibfnamefont [1]{#1}%
\providecommand \citenamefont [1]{#1}%
\providecommand \href@noop [0]{\@secondoftwo}%
\providecommand \href [0]{\begingroup \@sanitize@url \@href}%
\providecommand \@href[1]{\@@startlink{#1}\@@href}%
\providecommand \@@href[1]{\endgroup#1\@@endlink}%
\providecommand \@sanitize@url [0]{\catcode `\\12\catcode `\$12\catcode
  `\&12\catcode `\#12\catcode `\^12\catcode `\_12\catcode `\%12\relax}%
\providecommand \@@startlink[1]{}%
\providecommand \@@endlink[0]{}%
\providecommand \url  [0]{\begingroup\@sanitize@url \@url }%
\providecommand \@url [1]{\endgroup\@href {#1}{\urlprefix }}%
\providecommand \urlprefix  [0]{URL }%
\providecommand \Eprint [0]{\href }%
\providecommand \doibase [0]{https://doi.org/}%
\providecommand \selectlanguage [0]{\@gobble}%
\providecommand \bibinfo  [0]{\@secondoftwo}%
\providecommand \bibfield  [0]{\@secondoftwo}%
\providecommand \translation [1]{[#1]}%
\providecommand \BibitemOpen [0]{}%
\providecommand \bibitemStop [0]{}%
\providecommand \bibitemNoStop [0]{.\EOS\space}%
\providecommand \EOS [0]{\spacefactor3000\relax}%
\providecommand \BibitemShut  [1]{\csname bibitem#1\endcsname}%
\let\auto@bib@innerbib\@empty
%</preamble>
\bibitem [{\citenamefont {Tedrow}\ and\ \citenamefont
  {Meservey}(1973)}]{Tedrow73}%
  \BibitemOpen
  \bibfield  {author} {\bibinfo {author} {\bibfnamefont {P.~M.}\ \bibnamefont
  {Tedrow}}\ and\ \bibinfo {author} {\bibfnamefont {R.}~\bibnamefont
  {Meservey}},\ }\bibfield  {title} {\bibinfo {title} {Spin polarization of
  electrons tunneling from films of fe, co, ni, and gd},\ }\href
  {https://doi.org/10.1103/PhysRevB.7.318} {\bibfield  {journal} {\bibinfo
  {journal} {Phys. Rev. B}\ }\textbf {\bibinfo {volume} {7}},\ \bibinfo {pages}
  {318} (\bibinfo {year} {1973})}\BibitemShut {NoStop}%
\bibitem [{\citenamefont {Keizer}\ \emph {et~al.}(2006)\citenamefont {Keizer},
  \citenamefont {Goennenwein}, \citenamefont {Klapwijk}, \citenamefont {Miao},
  \citenamefont {Xiao},\ and\ \citenamefont {Gupta}}]{Keizer06}%
  \BibitemOpen
  \bibfield  {author} {\bibinfo {author} {\bibfnamefont {R.~S.}\ \bibnamefont
  {Keizer}}, \bibinfo {author} {\bibfnamefont {S.~T.~B.}\ \bibnamefont
  {Goennenwein}}, \bibinfo {author} {\bibfnamefont {T.~M.}\ \bibnamefont
  {Klapwijk}}, \bibinfo {author} {\bibfnamefont {G.}~\bibnamefont {Miao}},
  \bibinfo {author} {\bibfnamefont {G.}~\bibnamefont {Xiao}},\ and\ \bibinfo
  {author} {\bibfnamefont {A.}~\bibnamefont {Gupta}},\ }\bibfield  {title}
  {\bibinfo {title} {A spin triplet supercurrent through the half-metallic
  ferromagnet cro2},\ }\href {https://doi.org/10.1038/nature04499} {\bibfield
  {journal} {\bibinfo  {journal} {Nature}\ }\textbf {\bibinfo {volume} {439}},\
  \bibinfo {pages} {825} (\bibinfo {year} {2006})}\BibitemShut {NoStop}%
\bibitem [{\citenamefont {Khaire}\ \emph {et~al.}(2010)\citenamefont {Khaire},
  \citenamefont {Khasawneh}, \citenamefont {Pratt},\ and\ \citenamefont
  {Birge}}]{Khaire10}%
  \BibitemOpen
  \bibfield  {author} {\bibinfo {author} {\bibfnamefont {T.~S.}\ \bibnamefont
  {Khaire}}, \bibinfo {author} {\bibfnamefont {M.~A.}\ \bibnamefont
  {Khasawneh}}, \bibinfo {author} {\bibfnamefont {W.~P.}\ \bibnamefont
  {Pratt}},\ and\ \bibinfo {author} {\bibfnamefont {N.~O.}\ \bibnamefont
  {Birge}},\ }\bibfield  {title} {\bibinfo {title} {Observation of spin-triplet
  superconductivity in co-based josephson junctions},\ }\href
  {https://doi.org/10.1103/PhysRevLett.104.137002} {\bibfield  {journal}
  {\bibinfo  {journal} {Phys. Rev. Lett.}\ }\textbf {\bibinfo {volume} {104}},\
  \bibinfo {pages} {137002} (\bibinfo {year} {2010})}\BibitemShut {NoStop}%
\bibitem [{\citenamefont {Robinson}\ \emph {et~al.}(2010)\citenamefont
  {Robinson}, \citenamefont {Witt},\ and\ \citenamefont
  {Blamire}}]{Robinson10}%
  \BibitemOpen
  \bibfield  {author} {\bibinfo {author} {\bibfnamefont {J.~W.~A.}\
  \bibnamefont {Robinson}}, \bibinfo {author} {\bibfnamefont {J.~D.~S.}\
  \bibnamefont {Witt}},\ and\ \bibinfo {author} {\bibfnamefont {M.~G.}\
  \bibnamefont {Blamire}},\ }\bibfield  {title} {\bibinfo {title} {Controlled
  injection of spin-triplet supercurrents into a strong ferromagnet},\ }\href
  {https://doi.org/10.1126/science.1189246} {\bibfield  {journal} {\bibinfo
  {journal} {Science}\ }\textbf {\bibinfo {volume} {329}},\ \bibinfo {pages}
  {59} (\bibinfo {year} {2010})},\ \Eprint
  {https://arxiv.org/abs/https://www.science.org/doi/pdf/10.1126/science.1189246}
  {https://www.science.org/doi/pdf/10.1126/science.1189246} \BibitemShut
  {NoStop}%
\bibitem [{\citenamefont {Yang}\ \emph {et~al.}(2010)\citenamefont {Yang},
  \citenamefont {Yang}, \citenamefont {Takahashi}, \citenamefont {Maekawa},\
  and\ \citenamefont {Parkin}}]{YangSuper10}%
  \BibitemOpen
  \bibfield  {author} {\bibinfo {author} {\bibfnamefont {H.}~\bibnamefont
  {Yang}}, \bibinfo {author} {\bibfnamefont {S.-H.}\ \bibnamefont {Yang}},
  \bibinfo {author} {\bibfnamefont {S.}~\bibnamefont {Takahashi}}, \bibinfo
  {author} {\bibfnamefont {S.}~\bibnamefont {Maekawa}},\ and\ \bibinfo {author}
  {\bibfnamefont {S.~S.~P.}\ \bibnamefont {Parkin}},\ }\bibfield  {title}
  {\bibinfo {title} {Extremely long quasiparticle spin lifetimes in
  superconducting aluminium using mgo tunnel spin injectors},\ }\href
  {https://doi.org/10.1038/nmat2781} {\bibfield  {journal} {\bibinfo  {journal}
  {Nature Materials}\ }\textbf {\bibinfo {volume} {9}},\ \bibinfo {pages} {586}
  (\bibinfo {year} {2010})}\BibitemShut {NoStop}%
\bibitem [{\citenamefont {Jeon}\ \emph {et~al.}(2021)\citenamefont {Jeon},
  \citenamefont {Hazra}, \citenamefont {Cho}, \citenamefont {Chakraborty},
  \citenamefont {Jeon}, \citenamefont {Han}, \citenamefont {Meyerheim},
  \citenamefont {Kontos},\ and\ \citenamefont {Parkin}}]{Jeon21}%
  \BibitemOpen
  \bibfield  {author} {\bibinfo {author} {\bibfnamefont {K.-R.}\ \bibnamefont
  {Jeon}}, \bibinfo {author} {\bibfnamefont {B.~K.}\ \bibnamefont {Hazra}},
  \bibinfo {author} {\bibfnamefont {K.}~\bibnamefont {Cho}}, \bibinfo {author}
  {\bibfnamefont {A.}~\bibnamefont {Chakraborty}}, \bibinfo {author}
  {\bibfnamefont {J.-C.}\ \bibnamefont {Jeon}}, \bibinfo {author}
  {\bibfnamefont {H.}~\bibnamefont {Han}}, \bibinfo {author} {\bibfnamefont
  {H.~L.}\ \bibnamefont {Meyerheim}}, \bibinfo {author} {\bibfnamefont
  {T.}~\bibnamefont {Kontos}},\ and\ \bibinfo {author} {\bibfnamefont
  {S.~S.~P.}\ \bibnamefont {Parkin}},\ }\bibfield  {title} {\bibinfo {title}
  {Long-range supercurrents through a chiral non-collinear antiferromagnet in
  lateral josephson junctions},\ }\href
  {https://doi.org/10.1038/s41563-021-01061-9} {\bibfield  {journal} {\bibinfo
  {journal} {Nature Materials}\ }\textbf {\bibinfo {volume} {20}},\ \bibinfo
  {pages} {1358} (\bibinfo {year} {2021})}\BibitemShut {NoStop}%
\bibitem [{\citenamefont {Sanchez-Manzano}\ \emph {et~al.}(2022)\citenamefont
  {Sanchez-Manzano}, \citenamefont {Mesoraca}, \citenamefont {Cuellar},
  \citenamefont {Cabero}, \citenamefont {Rouco}, \citenamefont {Orfila},
  \citenamefont {Palermo}, \citenamefont {Balan}, \citenamefont {Marcano},
  \citenamefont {Sander}, \citenamefont {Rocci}, \citenamefont
  {Garcia-Barriocanal}, \citenamefont {Gallego}, \citenamefont {Tornos},
  \citenamefont {Rivera}, \citenamefont {Mompean}, \citenamefont
  {Garcia-Hernandez}, \citenamefont {Gonzalez-Calbet}, \citenamefont {Leon},
  \citenamefont {Valencia}, \citenamefont {Feuillet-Palma}, \citenamefont
  {Bergeal}, \citenamefont {Buzdin}, \citenamefont {Lesueur}, \citenamefont
  {Villegas},\ and\ \citenamefont {Santamaria}}]{Sanchez-Manzano22}%
  \BibitemOpen
  \bibfield  {author} {\bibinfo {author} {\bibfnamefont {D.}~\bibnamefont
  {Sanchez-Manzano}}, \bibinfo {author} {\bibfnamefont {S.}~\bibnamefont
  {Mesoraca}}, \bibinfo {author} {\bibfnamefont {F.~A.}\ \bibnamefont
  {Cuellar}}, \bibinfo {author} {\bibfnamefont {M.}~\bibnamefont {Cabero}},
  \bibinfo {author} {\bibfnamefont {V.}~\bibnamefont {Rouco}}, \bibinfo
  {author} {\bibfnamefont {G.}~\bibnamefont {Orfila}}, \bibinfo {author}
  {\bibfnamefont {X.}~\bibnamefont {Palermo}}, \bibinfo {author} {\bibfnamefont
  {A.}~\bibnamefont {Balan}}, \bibinfo {author} {\bibfnamefont
  {L.}~\bibnamefont {Marcano}}, \bibinfo {author} {\bibfnamefont
  {A.}~\bibnamefont {Sander}}, \bibinfo {author} {\bibfnamefont
  {M.}~\bibnamefont {Rocci}}, \bibinfo {author} {\bibfnamefont
  {J.}~\bibnamefont {Garcia-Barriocanal}}, \bibinfo {author} {\bibfnamefont
  {F.}~\bibnamefont {Gallego}}, \bibinfo {author} {\bibfnamefont
  {J.}~\bibnamefont {Tornos}}, \bibinfo {author} {\bibfnamefont
  {A.}~\bibnamefont {Rivera}}, \bibinfo {author} {\bibfnamefont
  {F.}~\bibnamefont {Mompean}}, \bibinfo {author} {\bibfnamefont
  {M.}~\bibnamefont {Garcia-Hernandez}}, \bibinfo {author} {\bibfnamefont
  {J.~M.}\ \bibnamefont {Gonzalez-Calbet}}, \bibinfo {author} {\bibfnamefont
  {C.}~\bibnamefont {Leon}}, \bibinfo {author} {\bibfnamefont {S.}~\bibnamefont
  {Valencia}}, \bibinfo {author} {\bibfnamefont {C.}~\bibnamefont
  {Feuillet-Palma}}, \bibinfo {author} {\bibfnamefont {N.}~\bibnamefont
  {Bergeal}}, \bibinfo {author} {\bibfnamefont {A.~I.}\ \bibnamefont {Buzdin}},
  \bibinfo {author} {\bibfnamefont {J.}~\bibnamefont {Lesueur}}, \bibinfo
  {author} {\bibfnamefont {J.~E.}\ \bibnamefont {Villegas}},\ and\ \bibinfo
  {author} {\bibfnamefont {J.}~\bibnamefont {Santamaria}},\ }\bibfield  {title}
  {\bibinfo {title} {Extremely long-range, high-temperature josephson coupling
  across a half-metallic ferromagnet},\ }\href
  {https://doi.org/10.1038/s41563-021-01162-5} {\bibfield  {journal} {\bibinfo
  {journal} {Nature Materials}\ }\textbf {\bibinfo {volume} {21}},\ \bibinfo
  {pages} {188} (\bibinfo {year} {2022})}\BibitemShut {NoStop}%
\bibitem [{\citenamefont {Ryazanov}\ \emph {et~al.}(2001)\citenamefont
  {Ryazanov}, \citenamefont {Oboznov}, \citenamefont {Rusanov}, \citenamefont
  {Veretennikov}, \citenamefont {Golubov},\ and\ \citenamefont
  {Aarts}}]{Ryazanov01}%
  \BibitemOpen
  \bibfield  {author} {\bibinfo {author} {\bibfnamefont {V.~V.}\ \bibnamefont
  {Ryazanov}}, \bibinfo {author} {\bibfnamefont {V.~A.}\ \bibnamefont
  {Oboznov}}, \bibinfo {author} {\bibfnamefont {A.~Y.}\ \bibnamefont
  {Rusanov}}, \bibinfo {author} {\bibfnamefont {A.~V.}\ \bibnamefont
  {Veretennikov}}, \bibinfo {author} {\bibfnamefont {A.~A.}\ \bibnamefont
  {Golubov}},\ and\ \bibinfo {author} {\bibfnamefont {J.}~\bibnamefont
  {Aarts}},\ }\bibfield  {title} {\bibinfo {title} {Coupling of two
  superconductors through a ferromagnet: Evidence for a $\ensuremath{\pi}$
  junction},\ }\href {https://doi.org/10.1103/PhysRevLett.86.2427} {\bibfield
  {journal} {\bibinfo  {journal} {Phys. Rev. Lett.}\ }\textbf {\bibinfo
  {volume} {86}},\ \bibinfo {pages} {2427} (\bibinfo {year}
  {2001})}\BibitemShut {NoStop}%
\bibitem [{\citenamefont {Jeon}\ \emph {et~al.}(2023)\citenamefont {Jeon},
  \citenamefont {Hazra}, \citenamefont {Kim}, \citenamefont {Jeon},
  \citenamefont {Han}, \citenamefont {Meyerheim}, \citenamefont {Kontos},
  \citenamefont {Cottet},\ and\ \citenamefont {Parkin}}]{Jeon23}%
  \BibitemOpen
  \bibfield  {author} {\bibinfo {author} {\bibfnamefont {K.-R.}\ \bibnamefont
  {Jeon}}, \bibinfo {author} {\bibfnamefont {B.~K.}\ \bibnamefont {Hazra}},
  \bibinfo {author} {\bibfnamefont {J.-K.}\ \bibnamefont {Kim}}, \bibinfo
  {author} {\bibfnamefont {J.-C.}\ \bibnamefont {Jeon}}, \bibinfo {author}
  {\bibfnamefont {H.}~\bibnamefont {Han}}, \bibinfo {author} {\bibfnamefont
  {H.~L.}\ \bibnamefont {Meyerheim}}, \bibinfo {author} {\bibfnamefont
  {T.}~\bibnamefont {Kontos}}, \bibinfo {author} {\bibfnamefont
  {A.}~\bibnamefont {Cottet}},\ and\ \bibinfo {author} {\bibfnamefont
  {S.~S.~P.}\ \bibnamefont {Parkin}},\ }\bibfield  {title} {\bibinfo {title}
  {Chiral antiferromagnetic josephson junctions as spin-triplet supercurrent
  spin valves and d.c. squids},\ }\href
  {https://doi.org/10.1038/s41565-023-01336-z} {\bibfield  {journal} {\bibinfo
  {journal} {Nature Nanotechnology}\ }\textbf {\bibinfo {volume} {18}},\
  \bibinfo {pages} {747} (\bibinfo {year} {2023})}\BibitemShut {NoStop}%
\bibitem [{\citenamefont {Jeon}\ \emph {et~al.}(2026)\citenamefont {Jeon},
  \citenamefont {Kim}, \citenamefont {Yoon}, \citenamefont {Jeon},
  \citenamefont {Han}, \citenamefont {Cottet}, \citenamefont {Kontos},\ and\
  \citenamefont {Parkin}}]{Jeon26}%
  \BibitemOpen
  \bibfield  {author} {\bibinfo {author} {\bibfnamefont {K.-R.}\ \bibnamefont
  {Jeon}}, \bibinfo {author} {\bibfnamefont {J.-K.}\ \bibnamefont {Kim}},
  \bibinfo {author} {\bibfnamefont {J.}~\bibnamefont {Yoon}}, \bibinfo {author}
  {\bibfnamefont {J.-C.}\ \bibnamefont {Jeon}}, \bibinfo {author}
  {\bibfnamefont {H.}~\bibnamefont {Han}}, \bibinfo {author} {\bibfnamefont
  {A.}~\bibnamefont {Cottet}}, \bibinfo {author} {\bibfnamefont
  {T.}~\bibnamefont {Kontos}},\ and\ \bibinfo {author} {\bibfnamefont
  {S.~S.~P.}\ \bibnamefont {Parkin}},\ }\bibfield  {title} {\bibinfo {title}
  {Interferometric evidence of nonvolatile anomalous phase shifts in
  exchange-spin-split josephson supercurrent diodes},\ }\href
  {https://doi.org/10.1021/acsnano.5c17979} {\bibfield  {journal} {\bibinfo
  {journal} {ACS Nano}\ }\textbf {\bibinfo {volume} {20}},\ \bibinfo {pages}
  {4384} (\bibinfo {year} {2026})}\BibitemShut {NoStop}%
\bibitem [{\citenamefont {Assouline}\ \emph {et~al.}(2019)\citenamefont
  {Assouline}, \citenamefont {Feuillet-Palma}, \citenamefont {Bergeal},
  \citenamefont {Zhang}, \citenamefont {Mottaghizadeh}, \citenamefont
  {Zimmers}, \citenamefont {Lhuillier}, \citenamefont {Eddrie}, \citenamefont
  {Atkinson}, \citenamefont {Aprili},\ and\ \citenamefont
  {Aubin}}]{Assouline19}%
  \BibitemOpen
  \bibfield  {author} {\bibinfo {author} {\bibfnamefont {A.}~\bibnamefont
  {Assouline}}, \bibinfo {author} {\bibfnamefont {C.}~\bibnamefont
  {Feuillet-Palma}}, \bibinfo {author} {\bibfnamefont {N.}~\bibnamefont
  {Bergeal}}, \bibinfo {author} {\bibfnamefont {T.}~\bibnamefont {Zhang}},
  \bibinfo {author} {\bibfnamefont {A.}~\bibnamefont {Mottaghizadeh}}, \bibinfo
  {author} {\bibfnamefont {A.}~\bibnamefont {Zimmers}}, \bibinfo {author}
  {\bibfnamefont {E.}~\bibnamefont {Lhuillier}}, \bibinfo {author}
  {\bibfnamefont {M.}~\bibnamefont {Eddrie}}, \bibinfo {author} {\bibfnamefont
  {P.}~\bibnamefont {Atkinson}}, \bibinfo {author} {\bibfnamefont
  {M.}~\bibnamefont {Aprili}},\ and\ \bibinfo {author} {\bibfnamefont
  {H.}~\bibnamefont {Aubin}},\ }\bibfield  {title} {\bibinfo {title}
  {Spin-orbit induced phase-shift in bi2se3 josephson junctions},\ }\href
  {https://doi.org/10.1038/s41467-018-08022-y} {\bibfield  {journal} {\bibinfo
  {journal} {Nature Communications}\ }\textbf {\bibinfo {volume} {10}},\
  \bibinfo {pages} {126} (\bibinfo {year} {2019})}\BibitemShut {NoStop}%
\bibitem [{\citenamefont {Pal}\ \emph {et~al.}(2022)\citenamefont {Pal},
  \citenamefont {Chakraborty}, \citenamefont {Sivakumar}, \citenamefont
  {Davydova}, \citenamefont {Gopi}, \citenamefont {Pandeya}, \citenamefont
  {Krieger}, \citenamefont {Zhang}, \citenamefont {Date}, \citenamefont {Ju},
  \citenamefont {Yuan}, \citenamefont {Schr{\"o}ter}, \citenamefont {Fu},\ and\
  \citenamefont {Parkin}}]{Pal22}%
  \BibitemOpen
  \bibfield  {author} {\bibinfo {author} {\bibfnamefont {B.}~\bibnamefont
  {Pal}}, \bibinfo {author} {\bibfnamefont {A.}~\bibnamefont {Chakraborty}},
  \bibinfo {author} {\bibfnamefont {P.~K.}\ \bibnamefont {Sivakumar}}, \bibinfo
  {author} {\bibfnamefont {M.}~\bibnamefont {Davydova}}, \bibinfo {author}
  {\bibfnamefont {A.~K.}\ \bibnamefont {Gopi}}, \bibinfo {author}
  {\bibfnamefont {A.~K.}\ \bibnamefont {Pandeya}}, \bibinfo {author}
  {\bibfnamefont {J.~A.}\ \bibnamefont {Krieger}}, \bibinfo {author}
  {\bibfnamefont {Y.}~\bibnamefont {Zhang}}, \bibinfo {author} {\bibfnamefont
  {M.}~\bibnamefont {Date}}, \bibinfo {author} {\bibfnamefont {S.}~\bibnamefont
  {Ju}}, \bibinfo {author} {\bibfnamefont {N.}~\bibnamefont {Yuan}}, \bibinfo
  {author} {\bibfnamefont {N.~B.~M.}\ \bibnamefont {Schr{\"o}ter}}, \bibinfo
  {author} {\bibfnamefont {L.}~\bibnamefont {Fu}},\ and\ \bibinfo {author}
  {\bibfnamefont {S.~S.~P.}\ \bibnamefont {Parkin}},\ }\bibfield  {title}
  {\bibinfo {title} {Josephson diode effect from cooper pair momentum in a
  topological semimetal},\ }\href {https://doi.org/10.1038/s41567-022-01699-5}
  {\bibfield  {journal} {\bibinfo  {journal} {Nature Physics}\ }\textbf
  {\bibinfo {volume} {18}},\ \bibinfo {pages} {1228} (\bibinfo {year}
  {2022})}\BibitemShut {NoStop}%
\bibitem [{\citenamefont {Zhang}\ \emph {et~al.}(2025)\citenamefont {Zhang},
  \citenamefont {Sun}, \citenamefont {Jia}, \citenamefont {Yang}, \citenamefont
  {Yan}, \citenamefont {Ai}, \citenamefont {Xie}, \citenamefont {Zhang},
  \citenamefont {Gao}, \citenamefont {Xu}, \citenamefont {Liu}, \citenamefont
  {Ma}, \citenamefont {Hu}, \citenamefont {Kou}, \citenamefont {Zou},
  \citenamefont {Ni}, \citenamefont {Law}, \citenamefont {Dong},\ and\
  \citenamefont {Xiu}}]{Zhang25}%
  \BibitemOpen
  \bibfield  {author} {\bibinfo {author} {\bibfnamefont {E.}~\bibnamefont
  {Zhang}}, \bibinfo {author} {\bibfnamefont {Z.-T.}\ \bibnamefont {Sun}},
  \bibinfo {author} {\bibfnamefont {Z.}~\bibnamefont {Jia}}, \bibinfo {author}
  {\bibfnamefont {J.}~\bibnamefont {Yang}}, \bibinfo {author} {\bibfnamefont
  {J.}~\bibnamefont {Yan}}, \bibinfo {author} {\bibfnamefont {L.}~\bibnamefont
  {Ai}}, \bibinfo {author} {\bibfnamefont {Y.-M.}\ \bibnamefont {Xie}},
  \bibinfo {author} {\bibfnamefont {Y.}~\bibnamefont {Zhang}}, \bibinfo
  {author} {\bibfnamefont {X.-J.}\ \bibnamefont {Gao}}, \bibinfo {author}
  {\bibfnamefont {X.}~\bibnamefont {Xu}}, \bibinfo {author} {\bibfnamefont
  {S.}~\bibnamefont {Liu}}, \bibinfo {author} {\bibfnamefont {Q.}~\bibnamefont
  {Ma}}, \bibinfo {author} {\bibfnamefont {C.}~\bibnamefont {Hu}}, \bibinfo
  {author} {\bibfnamefont {X.}~\bibnamefont {Kou}}, \bibinfo {author}
  {\bibfnamefont {J.}~\bibnamefont {Zou}}, \bibinfo {author} {\bibfnamefont
  {N.}~\bibnamefont {Ni}}, \bibinfo {author} {\bibfnamefont {K.~T.}\
  \bibnamefont {Law}}, \bibinfo {author} {\bibfnamefont {S.}~\bibnamefont
  {Dong}},\ and\ \bibinfo {author} {\bibfnamefont {F.}~\bibnamefont {Xiu}},\
  }\bibfield  {title} {\bibinfo {title} {Observation of edge supercurrent in
  topological antiferromagnet mnbi<sub>2</sub>te<sub>4</sub>-based josephson
  junctions},\ }\href {https://doi.org/10.1126/sciadv.ads8730} {\bibfield
  {journal} {\bibinfo  {journal} {Science Advances}\ }\textbf {\bibinfo
  {volume} {11}},\ \bibinfo {pages} {eads8730} (\bibinfo {year} {2025})},\
  \Eprint
  {https://arxiv.org/abs/https://www.science.org/doi/pdf/10.1126/sciadv.ads8730}
  {https://www.science.org/doi/pdf/10.1126/sciadv.ads8730} \BibitemShut
  {NoStop}%
\bibitem [{\citenamefont {Sato}\ and\ \citenamefont {Ando}(2017)}]{SatoAndo17}%
  \BibitemOpen
  \bibfield  {author} {\bibinfo {author} {\bibfnamefont {M.}~\bibnamefont
  {Sato}}\ and\ \bibinfo {author} {\bibfnamefont {Y.}~\bibnamefont {Ando}},\
  }\bibfield  {title} {\bibinfo {title} {Topological superconductors: a
  review},\ }\href {https://doi.org/10.1088/1361-6633/aa6ac7} {\bibfield
  {journal} {\bibinfo  {journal} {Reports on Progress in Physics}\ }\textbf
  {\bibinfo {volume} {80}},\ \bibinfo {pages} {076501} (\bibinfo {year}
  {2017})}\BibitemShut {NoStop}%
\bibitem [{\citenamefont {Amundsen}\ \emph {et~al.}(2024)\citenamefont
  {Amundsen}, \citenamefont {Linder}, \citenamefont {Robinson}, \citenamefont
  {\ifmmode \check{Z}\else \v{Z}\fi{}uti\ifmmode~\acute{c}\else \'{c}\fi{}},\
  and\ \citenamefont {Banerjee}}]{Amundsen24}%
  \BibitemOpen
  \bibfield  {author} {\bibinfo {author} {\bibfnamefont {M.}~\bibnamefont
  {Amundsen}}, \bibinfo {author} {\bibfnamefont {J.}~\bibnamefont {Linder}},
  \bibinfo {author} {\bibfnamefont {J.~W.~A.}\ \bibnamefont {Robinson}},
  \bibinfo {author} {\bibfnamefont {I.}~\bibnamefont {\ifmmode \check{Z}\else
  \v{Z}\fi{}uti\ifmmode~\acute{c}\else \'{c}\fi{}}},\ and\ \bibinfo {author}
  {\bibfnamefont {N.}~\bibnamefont {Banerjee}},\ }\bibfield  {title} {\bibinfo
  {title} {Colloquium: Spin-orbit effects in superconducting hybrid
  structures},\ }\href {https://doi.org/10.1103/RevModPhys.96.021003}
  {\bibfield  {journal} {\bibinfo  {journal} {Rev. Mod. Phys.}\ }\textbf
  {\bibinfo {volume} {96}},\ \bibinfo {pages} {021003} (\bibinfo {year}
  {2024})}\BibitemShut {NoStop}%
\bibitem [{\citenamefont {Bergeret}\ and\ \citenamefont
  {Tokatly}(2016)}]{Bergeret16}%
  \BibitemOpen
  \bibfield  {author} {\bibinfo {author} {\bibfnamefont {F.~S.}\ \bibnamefont
  {Bergeret}}\ and\ \bibinfo {author} {\bibfnamefont {I.~V.}\ \bibnamefont
  {Tokatly}},\ }\bibfield  {title} {\bibinfo {title} {Manifestation of
  extrinsic spin hall effect in superconducting structures: Nondissipative
  magnetoelectric effects},\ }\href
  {https://doi.org/10.1103/PhysRevB.94.180502} {\bibfield  {journal} {\bibinfo
  {journal} {Phys. Rev. B}\ }\textbf {\bibinfo {volume} {94}},\ \bibinfo
  {pages} {180502} (\bibinfo {year} {2016})}\BibitemShut {NoStop}%
\bibitem [{\citenamefont {Daido}\ \emph {et~al.}(2022)\citenamefont {Daido},
  \citenamefont {Ikeda},\ and\ \citenamefont {Yanase}}]{Daido22}%
  \BibitemOpen
  \bibfield  {author} {\bibinfo {author} {\bibfnamefont {A.}~\bibnamefont
  {Daido}}, \bibinfo {author} {\bibfnamefont {Y.}~\bibnamefont {Ikeda}},\ and\
  \bibinfo {author} {\bibfnamefont {Y.}~\bibnamefont {Yanase}},\ }\bibfield
  {title} {\bibinfo {title} {Intrinsic superconducting diode effect},\ }\href
  {https://doi.org/10.1103/PhysRevLett.128.037001} {\bibfield  {journal}
  {\bibinfo  {journal} {Phys. Rev. Lett.}\ }\textbf {\bibinfo {volume} {128}},\
  \bibinfo {pages} {037001} (\bibinfo {year} {2022})}\BibitemShut {NoStop}%
\bibitem [{\citenamefont {He}\ \emph {et~al.}(2022)\citenamefont {He},
  \citenamefont {Tanaka},\ and\ \citenamefont {Nagaosa}}]{He22}%
  \BibitemOpen
  \bibfield  {author} {\bibinfo {author} {\bibfnamefont {J.~J.}\ \bibnamefont
  {He}}, \bibinfo {author} {\bibfnamefont {Y.}~\bibnamefont {Tanaka}},\ and\
  \bibinfo {author} {\bibfnamefont {N.}~\bibnamefont {Nagaosa}},\ }\bibfield
  {title} {\bibinfo {title} {A phenomenological theory of superconductor
  diodes},\ }\href {https://doi.org/10.1088/1367-2630/ac6766} {\bibfield
  {journal} {\bibinfo  {journal} {New Journal of Physics}\ }\textbf {\bibinfo
  {volume} {24}},\ \bibinfo {pages} {053014} (\bibinfo {year}
  {2022})}\BibitemShut {NoStop}%
\bibitem [{\citenamefont {Cayao}\ \emph {et~al.}(2024)\citenamefont {Cayao},
  \citenamefont {Nagaosa},\ and\ \citenamefont {Tanaka}}]{Cayao24}%
  \BibitemOpen
  \bibfield  {author} {\bibinfo {author} {\bibfnamefont {J.}~\bibnamefont
  {Cayao}}, \bibinfo {author} {\bibfnamefont {N.}~\bibnamefont {Nagaosa}},\
  and\ \bibinfo {author} {\bibfnamefont {Y.}~\bibnamefont {Tanaka}},\
  }\bibfield  {title} {\bibinfo {title} {Enhancing the josephson diode effect
  with majorana bound states},\ }\href
  {https://doi.org/10.1103/PhysRevB.109.L081405} {\bibfield  {journal}
  {\bibinfo  {journal} {Phys. Rev. B}\ }\textbf {\bibinfo {volume} {109}},\
  \bibinfo {pages} {L081405} (\bibinfo {year} {2024})}\BibitemShut {NoStop}%
\bibitem [{\citenamefont {Kokkeler}\ \emph {et~al.}(2024)\citenamefont
  {Kokkeler}, \citenamefont {Tokatly},\ and\ \citenamefont
  {Bergeret}}]{Kokkeler24}%
  \BibitemOpen
  \bibfield  {author} {\bibinfo {author} {\bibfnamefont {T.}~\bibnamefont
  {Kokkeler}}, \bibinfo {author} {\bibfnamefont {I.}~\bibnamefont {Tokatly}},\
  and\ \bibinfo {author} {\bibfnamefont {F.~S.}\ \bibnamefont {Bergeret}},\
  }\bibfield  {title} {\bibinfo {title} {{Nonreciprocal superconducting
  transport and the spin Hall effect in gyrotropic structures}},\ }\href
  {https://doi.org/10.21468/SciPostPhys.16.2.055} {\bibfield  {journal}
  {\bibinfo  {journal} {SciPost Phys.}\ }\textbf {\bibinfo {volume} {16}},\
  \bibinfo {pages} {055} (\bibinfo {year} {2024})}\BibitemShut {NoStop}%
\bibitem [{\citenamefont {Yerin}\ \emph {et~al.}(2024)\citenamefont {Yerin},
  \citenamefont {Drechsler}, \citenamefont {Varlamov}, \citenamefont {Cuoco},\
  and\ \citenamefont {Giazotto}}]{Yerin24}%
  \BibitemOpen
  \bibfield  {author} {\bibinfo {author} {\bibfnamefont {Y.}~\bibnamefont
  {Yerin}}, \bibinfo {author} {\bibfnamefont {S.-L.}\ \bibnamefont
  {Drechsler}}, \bibinfo {author} {\bibfnamefont {A.~A.}\ \bibnamefont
  {Varlamov}}, \bibinfo {author} {\bibfnamefont {M.}~\bibnamefont {Cuoco}},\
  and\ \bibinfo {author} {\bibfnamefont {F.}~\bibnamefont {Giazotto}},\
  }\bibfield  {title} {\bibinfo {title} {Supercurrent rectification with
  time-reversal symmetry broken multiband superconductors},\ }\href
  {https://doi.org/10.1103/PhysRevB.110.054501} {\bibfield  {journal} {\bibinfo
   {journal} {Phys. Rev. B}\ }\textbf {\bibinfo {volume} {110}},\ \bibinfo
  {pages} {054501} (\bibinfo {year} {2024})}\BibitemShut {NoStop}%
\bibitem [{\citenamefont {Sahoo}\ and\ \citenamefont {Soori}(2026)}]{Sahoo26}%
  \BibitemOpen
  \bibfield  {author} {\bibinfo {author} {\bibfnamefont {B.~K.}\ \bibnamefont
  {Sahoo}}\ and\ \bibinfo {author} {\bibfnamefont {A.}~\bibnamefont {Soori}},\
  }\href {https://arxiv.org/abs/2509.14109} {\bibinfo {title} {Giant field-free
  transverse josephson diode effect in altermagnets}} (\bibinfo {year}
  {2026}),\ \Eprint {https://arxiv.org/abs/2509.14109} {arXiv:2509.14109
  [cond-mat.mes-hall]} \BibitemShut {NoStop}%
\bibitem [{\citenamefont {Ando}\ \emph {et~al.}(2020)\citenamefont {Ando},
  \citenamefont {Miyasaka}, \citenamefont {Li}, \citenamefont {Ishizuka},
  \citenamefont {Arakawa}, \citenamefont {Shiota}, \citenamefont {Moriyama},
  \citenamefont {Yanase},\ and\ \citenamefont {Ono}}]{Ando20}%
  \BibitemOpen
  \bibfield  {author} {\bibinfo {author} {\bibfnamefont {F.}~\bibnamefont
  {Ando}}, \bibinfo {author} {\bibfnamefont {Y.}~\bibnamefont {Miyasaka}},
  \bibinfo {author} {\bibfnamefont {T.}~\bibnamefont {Li}}, \bibinfo {author}
  {\bibfnamefont {J.}~\bibnamefont {Ishizuka}}, \bibinfo {author}
  {\bibfnamefont {T.}~\bibnamefont {Arakawa}}, \bibinfo {author} {\bibfnamefont
  {Y.}~\bibnamefont {Shiota}}, \bibinfo {author} {\bibfnamefont
  {T.}~\bibnamefont {Moriyama}}, \bibinfo {author} {\bibfnamefont
  {Y.}~\bibnamefont {Yanase}},\ and\ \bibinfo {author} {\bibfnamefont
  {T.}~\bibnamefont {Ono}},\ }\bibfield  {title} {\bibinfo {title} {Observation
  of superconducting diode effect},\ }\href
  {https://doi.org/10.1038/s41586-020-2590-4} {\bibfield  {journal} {\bibinfo
  {journal} {Nature}\ }\textbf {\bibinfo {volume} {584}},\ \bibinfo {pages}
  {373} (\bibinfo {year} {2020})}\BibitemShut {NoStop}%
\bibitem [{\citenamefont {Baumgartner}\ \emph {et~al.}(2022)\citenamefont
  {Baumgartner}, \citenamefont {Fuchs}, \citenamefont {Costa}, \citenamefont
  {Reinhardt}, \citenamefont {Gronin}, \citenamefont {Gardner}, \citenamefont
  {Lindemann}, \citenamefont {Manfra}, \citenamefont {Faria~Junior},
  \citenamefont {Kochan}, \citenamefont {Fabian}, \citenamefont {Paradiso},\
  and\ \citenamefont {Strunk}}]{Baumgartner22}%
  \BibitemOpen
  \bibfield  {author} {\bibinfo {author} {\bibfnamefont {C.}~\bibnamefont
  {Baumgartner}}, \bibinfo {author} {\bibfnamefont {L.}~\bibnamefont {Fuchs}},
  \bibinfo {author} {\bibfnamefont {A.}~\bibnamefont {Costa}}, \bibinfo
  {author} {\bibfnamefont {S.}~\bibnamefont {Reinhardt}}, \bibinfo {author}
  {\bibfnamefont {S.}~\bibnamefont {Gronin}}, \bibinfo {author} {\bibfnamefont
  {G.~C.}\ \bibnamefont {Gardner}}, \bibinfo {author} {\bibfnamefont
  {T.}~\bibnamefont {Lindemann}}, \bibinfo {author} {\bibfnamefont {M.~J.}\
  \bibnamefont {Manfra}}, \bibinfo {author} {\bibfnamefont {P.~E.}\
  \bibnamefont {Faria~Junior}}, \bibinfo {author} {\bibfnamefont
  {D.}~\bibnamefont {Kochan}}, \bibinfo {author} {\bibfnamefont
  {J.}~\bibnamefont {Fabian}}, \bibinfo {author} {\bibfnamefont
  {N.}~\bibnamefont {Paradiso}},\ and\ \bibinfo {author} {\bibfnamefont
  {C.}~\bibnamefont {Strunk}},\ }\bibfield  {title} {\bibinfo {title}
  {Supercurrent rectification and magnetochiral effects in symmetric josephson
  junctions},\ }\href {https://doi.org/10.1038/s41565-021-01009-9} {\bibfield
  {journal} {\bibinfo  {journal} {Nature Nanotechnology}\ }\textbf {\bibinfo
  {volume} {17}},\ \bibinfo {pages} {39} (\bibinfo {year} {2022})}\BibitemShut
  {NoStop}%
\bibitem [{\citenamefont {Jeon}\ \emph {et~al.}(2022)\citenamefont {Jeon},
  \citenamefont {Kim}, \citenamefont {Yoon}, \citenamefont {Jeon},
  \citenamefont {Han}, \citenamefont {Cottet}, \citenamefont {Kontos},\ and\
  \citenamefont {Parkin}}]{Jeon22}%
  \BibitemOpen
  \bibfield  {author} {\bibinfo {author} {\bibfnamefont {K.-R.}\ \bibnamefont
  {Jeon}}, \bibinfo {author} {\bibfnamefont {J.-K.}\ \bibnamefont {Kim}},
  \bibinfo {author} {\bibfnamefont {J.}~\bibnamefont {Yoon}}, \bibinfo {author}
  {\bibfnamefont {J.-C.}\ \bibnamefont {Jeon}}, \bibinfo {author}
  {\bibfnamefont {H.}~\bibnamefont {Han}}, \bibinfo {author} {\bibfnamefont
  {A.}~\bibnamefont {Cottet}}, \bibinfo {author} {\bibfnamefont
  {T.}~\bibnamefont {Kontos}},\ and\ \bibinfo {author} {\bibfnamefont
  {S.~S.~P.}\ \bibnamefont {Parkin}},\ }\bibfield  {title} {\bibinfo {title}
  {Zero-field polarity-reversible josephson supercurrent diodes enabled by a
  proximity-magnetized pt barrier},\ }\href
  {https://doi.org/10.1038/s41563-022-01300-7} {\bibfield  {journal} {\bibinfo
  {journal} {Nature Materials}\ }\textbf {\bibinfo {volume} {21}},\ \bibinfo
  {pages} {1008} (\bibinfo {year} {2022})}\BibitemShut {NoStop}%
\bibitem [{\citenamefont {Reinhardt}\ \emph {et~al.}(2024)\citenamefont
  {Reinhardt}, \citenamefont {Ascherl}, \citenamefont {Costa}, \citenamefont
  {Berger}, \citenamefont {Gronin}, \citenamefont {Gardner}, \citenamefont
  {Lindemann}, \citenamefont {Manfra}, \citenamefont {Fabian}, \citenamefont
  {Kochan}, \citenamefont {Strunk},\ and\ \citenamefont
  {Paradiso}}]{Reinhardt24}%
  \BibitemOpen
  \bibfield  {author} {\bibinfo {author} {\bibfnamefont {S.}~\bibnamefont
  {Reinhardt}}, \bibinfo {author} {\bibfnamefont {T.}~\bibnamefont {Ascherl}},
  \bibinfo {author} {\bibfnamefont {A.}~\bibnamefont {Costa}}, \bibinfo
  {author} {\bibfnamefont {J.}~\bibnamefont {Berger}}, \bibinfo {author}
  {\bibfnamefont {S.}~\bibnamefont {Gronin}}, \bibinfo {author} {\bibfnamefont
  {G.~C.}\ \bibnamefont {Gardner}}, \bibinfo {author} {\bibfnamefont
  {T.}~\bibnamefont {Lindemann}}, \bibinfo {author} {\bibfnamefont {M.~J.}\
  \bibnamefont {Manfra}}, \bibinfo {author} {\bibfnamefont {J.}~\bibnamefont
  {Fabian}}, \bibinfo {author} {\bibfnamefont {D.}~\bibnamefont {Kochan}},
  \bibinfo {author} {\bibfnamefont {C.}~\bibnamefont {Strunk}},\ and\ \bibinfo
  {author} {\bibfnamefont {N.}~\bibnamefont {Paradiso}},\ }\bibfield  {title}
  {\bibinfo {title} {Link between supercurrent diode and anomalous josephson
  effect revealed by gate-controlled interferometry},\ }\href
  {https://doi.org/10.1038/s41467-024-48741-z} {\bibfield  {journal} {\bibinfo
  {journal} {Nature Communications}\ }\textbf {\bibinfo {volume} {15}},\
  \bibinfo {pages} {4413} (\bibinfo {year} {2024})}\BibitemShut {NoStop}%
\bibitem [{\citenamefont {Edelstein}(1990)}]{Edelstein90}%
  \BibitemOpen
  \bibfield  {author} {\bibinfo {author} {\bibfnamefont {V.}~\bibnamefont
  {Edelstein}},\ }\bibfield  {title} {\bibinfo {title} {Spin polarization of
  conduction electrons induced by electric current in two-dimensional
  asymmetric electron systems},\ }\href
  {https://doi.org/10.1016/0038-1098(90)90963-C} {\bibfield  {journal}
  {\bibinfo  {journal} {Solid State Communications}\ }\textbf {\bibinfo
  {volume} {73}},\ \bibinfo {pages} {233 } (\bibinfo {year}
  {1990})}\BibitemShut {NoStop}%
\bibitem [{\citenamefont {Edelstein}(1995)}]{Edelstein95}%
  \BibitemOpen
  \bibfield  {author} {\bibinfo {author} {\bibfnamefont {V.~M.}\ \bibnamefont
  {Edelstein}},\ }\bibfield  {title} {\bibinfo {title} {Magnetoelectric effect
  in polar superconductors},\ }\href
  {https://doi.org/10.1103/PhysRevLett.75.2004} {\bibfield  {journal} {\bibinfo
   {journal} {Phys. Rev. Lett.}\ }\textbf {\bibinfo {volume} {75}},\ \bibinfo
  {pages} {2004} (\bibinfo {year} {1995})}\BibitemShut {NoStop}%
\bibitem [{\citenamefont {Konschelle}\ \emph {et~al.}(2015)\citenamefont
  {Konschelle}, \citenamefont {Tokatly},\ and\ \citenamefont
  {Bergeret}}]{Konschelle15}%
  \BibitemOpen
  \bibfield  {author} {\bibinfo {author} {\bibfnamefont {F.~m.~c.}\
  \bibnamefont {Konschelle}}, \bibinfo {author} {\bibfnamefont {I.~V.}\
  \bibnamefont {Tokatly}},\ and\ \bibinfo {author} {\bibfnamefont {F.~S.}\
  \bibnamefont {Bergeret}},\ }\bibfield  {title} {\bibinfo {title} {Theory of
  the spin-galvanic effect and the anomalous phase shift
  ${\ensuremath{\varphi}}_{0}$ in superconductors and josephson junctions with
  intrinsic spin-orbit coupling},\ }\href
  {https://doi.org/10.1103/PhysRevB.92.125443} {\bibfield  {journal} {\bibinfo
  {journal} {Phys. Rev. B}\ }\textbf {\bibinfo {volume} {92}},\ \bibinfo
  {pages} {125443} (\bibinfo {year} {2015})}\BibitemShut {NoStop}%
\bibitem [{\citenamefont {Tatara}(2019)}]{TataraReview19}%
  \BibitemOpen
  \bibfield  {author} {\bibinfo {author} {\bibfnamefont {G.}~\bibnamefont
  {Tatara}},\ }\bibfield  {title} {\bibinfo {title} {Effective gauge field
  theory of spintronics},\ }\href
  {https://doi.org/https://doi.org/10.1016/j.physe.2018.05.011} {\bibfield
  {journal} {\bibinfo  {journal} {Physica E: Low-dimensional Systems and
  Nanostructures}\ }\textbf {\bibinfo {volume} {106}},\ \bibinfo {pages} {208 }
  (\bibinfo {year} {2019})}\BibitemShut {NoStop}%
\bibitem [{\citenamefont {Hirsch}(1999)}]{Hirsch99}%
  \BibitemOpen
  \bibfield  {author} {\bibinfo {author} {\bibfnamefont {J.~E.}\ \bibnamefont
  {Hirsch}},\ }\bibfield  {title} {\bibinfo {title} {Spin hall effect},\ }\href
  {https://doi.org/10.1103/PhysRevLett.83.1834} {\bibfield  {journal} {\bibinfo
   {journal} {Phys. Rev. Lett.}\ }\textbf {\bibinfo {volume} {83}},\ \bibinfo
  {pages} {1834} (\bibinfo {year} {1999})}\BibitemShut {NoStop}%
\bibitem [{\citenamefont {Tatara}(2018)}]{TataraSH18}%
  \BibitemOpen
  \bibfield  {author} {\bibinfo {author} {\bibfnamefont {G.}~\bibnamefont
  {Tatara}},\ }\bibfield  {title} {\bibinfo {title} {Spin correlation function
  theory of spin-charge conversion effects},\ }\href
  {https://doi.org/10.1103/PhysRevB.98.174422} {\bibfield  {journal} {\bibinfo
  {journal} {Phys. Rev. B}\ }\textbf {\bibinfo {volume} {98}},\ \bibinfo
  {pages} {174422} (\bibinfo {year} {2018})}\BibitemShut {NoStop}%
\bibitem [{\citenamefont {Golubov}\ \emph {et~al.}(2004)\citenamefont
  {Golubov}, \citenamefont {Kupriyanov},\ and\ \citenamefont
  {Il'ichev}}]{Golubov04}%
  \BibitemOpen
  \bibfield  {author} {\bibinfo {author} {\bibfnamefont {A.~A.}\ \bibnamefont
  {Golubov}}, \bibinfo {author} {\bibfnamefont {M.~Y.}\ \bibnamefont
  {Kupriyanov}},\ and\ \bibinfo {author} {\bibfnamefont {E.}~\bibnamefont
  {Il'ichev}},\ }\bibfield  {title} {\bibinfo {title} {The current-phase
  relation in josephson junctions},\ }\href
  {https://doi.org/10.1103/RevModPhys.76.411} {\bibfield  {journal} {\bibinfo
  {journal} {Rev. Mod. Phys.}\ }\textbf {\bibinfo {volume} {76}},\ \bibinfo
  {pages} {411} (\bibinfo {year} {2004})}\BibitemShut {NoStop}%
\bibitem [{\citenamefont {Kresin}(1986)}]{Kresin86}%
  \BibitemOpen
  \bibfield  {author} {\bibinfo {author} {\bibfnamefont {V.~Z.}\ \bibnamefont
  {Kresin}},\ }\bibfield  {title} {\bibinfo {title} {Josephson current in
  low-dimensional proximity systems and the field effect},\ }\href
  {https://doi.org/10.1103/PhysRevB.34.7587} {\bibfield  {journal} {\bibinfo
  {journal} {Phys. Rev. B}\ }\textbf {\bibinfo {volume} {34}},\ \bibinfo
  {pages} {7587} (\bibinfo {year} {1986})}\BibitemShut {NoStop}%
\bibitem [{\citenamefont {Coleman}(2015)}]{ColemanIMBP15}%
  \BibitemOpen
  \bibfield  {author} {\bibinfo {author} {\bibfnamefont {P.}~\bibnamefont
  {Coleman}},\ }\href
  {http://gen.lib.rus.ec/book/index.php?md5=a826f7c5d77f0b9afb3806e985635904}
  {\emph {\bibinfo {title} {Introduction to many-body physics}}}\ (\bibinfo
  {publisher} {Cambridge University Press},\ \bibinfo {year}
  {2015})\BibitemShut {NoStop}%
\bibitem [{\citenamefont {Dugaev}\ \emph {et~al.}(2001)\citenamefont {Dugaev},
  \citenamefont {Cr\'epieux},\ and\ \citenamefont {Bruno}}]{Dugaev01}%
  \BibitemOpen
  \bibfield  {author} {\bibinfo {author} {\bibfnamefont {V.~K.}\ \bibnamefont
  {Dugaev}}, \bibinfo {author} {\bibfnamefont {A.}~\bibnamefont {Cr\'epieux}},\
  and\ \bibinfo {author} {\bibfnamefont {P.}~\bibnamefont {Bruno}},\ }\bibfield
   {title} {\bibinfo {title} {Localization corrections to the anomalous hall
  effect in a ferromagnet},\ }\href
  {https://doi.org/10.1103/PhysRevB.64.104411} {\bibfield  {journal} {\bibinfo
  {journal} {Phys. Rev. B}\ }\textbf {\bibinfo {volume} {64}},\ \bibinfo
  {pages} {104411} (\bibinfo {year} {2001})}\BibitemShut {NoStop}%
\bibitem [{\citenamefont {Shitade}\ and\ \citenamefont
  {Tatara}(2022)}]{Shitade22}%
  \BibitemOpen
  \bibfield  {author} {\bibinfo {author} {\bibfnamefont {A.}~\bibnamefont
  {Shitade}}\ and\ \bibinfo {author} {\bibfnamefont {G.}~\bibnamefont
  {Tatara}},\ }\bibfield  {title} {\bibinfo {title} {Spin accumulation without
  spin current},\ }\href {https://doi.org/10.1103/PhysRevB.105.L201202}
  {\bibfield  {journal} {\bibinfo  {journal} {Phys. Rev. B}\ }\textbf {\bibinfo
  {volume} {105}},\ \bibinfo {pages} {L201202} (\bibinfo {year}
  {2022})}\BibitemShut {NoStop}%
\bibitem [{\citenamefont {Matsuo}\ \emph {et~al.}(2023)\citenamefont {Matsuo},
  \citenamefont {Imoto}, \citenamefont {Yokoyama}, \citenamefont {Sato},
  \citenamefont {Lindemann}, \citenamefont {Gronin}, \citenamefont {Gardner},
  \citenamefont {Manfra},\ and\ \citenamefont {Tarucha}}]{MatsuoSA23}%
  \BibitemOpen
  \bibfield  {author} {\bibinfo {author} {\bibfnamefont {S.}~\bibnamefont
  {Matsuo}}, \bibinfo {author} {\bibfnamefont {T.}~\bibnamefont {Imoto}},
  \bibinfo {author} {\bibfnamefont {T.}~\bibnamefont {Yokoyama}}, \bibinfo
  {author} {\bibfnamefont {Y.}~\bibnamefont {Sato}}, \bibinfo {author}
  {\bibfnamefont {T.}~\bibnamefont {Lindemann}}, \bibinfo {author}
  {\bibfnamefont {S.}~\bibnamefont {Gronin}}, \bibinfo {author} {\bibfnamefont
  {G.~C.}\ \bibnamefont {Gardner}}, \bibinfo {author} {\bibfnamefont {M.~J.}\
  \bibnamefont {Manfra}},\ and\ \bibinfo {author} {\bibfnamefont
  {S.}~\bibnamefont {Tarucha}},\ }\bibfield  {title} {\bibinfo {title} {Phase
  engineering of anomalous josephson effect derived from andreev molecules},\
  }\href {https://doi.org/10.1126/sciadv.adj3698} {\bibfield  {journal}
  {\bibinfo  {journal} {Science Advances}\ }\textbf {\bibinfo {volume} {9}},\
  \bibinfo {pages} {eadj3698} (\bibinfo {year} {2023})},\ \Eprint
  {https://arxiv.org/abs/https://www.science.org/doi/pdf/10.1126/sciadv.adj3698}
  {https://www.science.org/doi/pdf/10.1126/sciadv.adj3698} \BibitemShut
  {NoStop}%
\end{thebibliography}%
\end{document}